\documentclass[11pt,a4paper]{article}

\usepackage[dvipsnames]{xcolor}
\usepackage[skip=1pt,labelfont=bf]{caption}
\usepackage{bm}
\usepackage{tikz}
\usepackage{graphicx}
\usepackage{braket}
\usepackage{cancel}
\usepackage{bbm}
\usepackage{mathtools}
\usepackage{amsmath}				% collection de symboles mathmatiques
\usepackage{amssymb}				% collection de symboles mathmatiques
\usepackage[latin1]{inputenc}      % utilisation directe des caractres accentus sur pc
\usepackage[T1]{fontenc}			% codage moderne des caractres sous Latex
\usepackage{stmaryrd}
\usepackage{comment}
\usepackage[colorlinks=true,breaklinks=true,allcolors=blue]{hyperref}
\usepackage[capitalize]{cleveref}
\usepackage{authblk}

\newcommand\bbone{\ensuremath{\mathbbm{1}}} 
\newcommand\numberthis{\addtocounter{equation}{1}\tag{\theequation}}

\newtheorem{definition}{Definition} 

\DeclareMathOperator{\sgn}{sgn}
\DeclareMathOperator{\Tr}{Tr}
\DeclareMathOperator{\Span}{Span}
\DeclareMathOperator{\myIm}{Im}

\allowdisplaybreaks

\begin{document}

\title{Quantum kinetic perturbation theory\\ for near-integrable spin chains with weak long-range interactions}

\author[1,2,3]{Cl\'ement Duval}
\author[3,4]{Michael Kastner}
\affil[1]{\small Universit\'e de Lyon, {\'E}cole Normale Sup\'erieure de Lyon, 46 All\'ee d'Italie, 69364 Lyon cedex 07, France}
\affil[2]{Universit\'e Paris Diderot, 75013 Paris, France}
\affil[3]{National Institute for Theoretical Physics (NITheP), Stellenbosch 7600, South Africa}
\affil[4]{\mbox{Institute of Theoretical Physics, Department of Physics, Stellenbosch University, Stellenbosch 7600, South Africa}}

\maketitle

\begin{abstract}
For a transverse-field Ising chain with weak long-range interactions we develop a perturbative scheme, based on quantum kinetic equations, around the integrable nearest-neighbour model. We introduce, discuss, and benchmark several truncations of the time evolution equations up to eighth order in the Jordan-Wigner fermionic operators. The resulting set of differential equations can be solved for lattices with $O(10^2)$ sites and facilitates the computation of spin expectation values and correlation functions to high accuracy, at least for moderate timescales. We use this scheme to study the relaxation dynamics of the model, involving prethermalisation and thermalisation. The techniques developed here can be generalised to other spin models with weak integrability-breaking terms.
\end{abstract}

%\tableofcontents

\section{Introduction}

Equilibration and thermalisation are topics that link nonequilibrium physics to equilibrium physics, and they play a fundamental role for the validity and success of thermodynamics. These topics have a long history and have been studied in a variety of settings, including classical mechanics {\em vs.}\ quantum mechanics, closed systems {\em vs.}\ open systems, and others. Renewed interest in equilibration and thermalisation in isolated quantum systems was to a large extend triggered by experimental progress in preparing and manipulating assemblies of cold atoms that are extremely well isolated from their surroundings; see \cite{EisertFriesdorfGogolin15,GogolinEisert16} for reviews. Near-integrable systems, consisting of a dominant integrable part plus a small integrability-breaking perturbation, have been studied early on in some of these experiments, including the celebrated quantum Newton's cradle by Kinoshita {\em et al.}\ \cite{Kinoshita_etal06}. The integrability-breaking perturbation ensures thermalisation to a microcanonical equilibrium, and the relaxation dynamics towards equilibrium in a near-integrable system takes place in two stages on widely separated timescales \cite{LangenGasenzerSchmiedmayer16,Mori_etal18,Tang_etal18,ReimannDabelow19}: A fast decay, termed {\em prethermalisation}, to a long-lasting nonequilibrium state that is characterised by a so-called {\em generalised Gibbs ensemble}\/ (GGE) \cite{Rigol_etal07,VidmarRigol16}; and a second step, in which relaxation to thermal equilibrium, as described by the ordinary Gibbs ensemble, occurs on a much longer timescale, once the integrability-breaking perturbation becomes relevant. 

Accurate and reliable calculations of these phenomena are challenging, at least when going beyond the small system sizes of $O(10)$ where exact diagonalisation is feasible. Perturbative techniques around the integrable limit suggest themselves for the problem at hand, and various types of such techniques have been employed in the context of prethermalisation, including a flow-equation methods \cite{MoeckelKehrein08}, self-consistent mean-field techniques \cite{BertiniFagotti15}, and quantum kinetic theory \cite{FuerstMendlSpohn13,TavoraMitra13,Bertini_etal15,Bertini_etal16}. The notion of {\em quantum kinetic theory}\/ subsumes a number of approximate methods based on identifying certain classes of operators (usually those of higher degree in the normal-ordered ladder operators; see \cref{Selection_truncation} for more precise statements) as negligible, and deriving a reduced set of equations of motion for the remaining operators only \cite{Bonitz}. In the abovementioned Refs.~\cite{FuerstMendlSpohn13,TavoraMitra13,Bertini_etal15,Bertini_etal16} quantum kinetic theories are developed for studying bosons or fermions in one spatial dimension.

The quantum kinetic theory we develop in the present paper differs from those works in several important aspects. The integrable part of the Hamiltonian $\mathcal{H}_\text{int}$ we consider is an Ising spin chain with nearest-neighbour interaction and a transverse magnetic field, see Eq.~\eqref{Hint}. Our aim is to study the effect of weak long-range interactions, where we define {\em long range}\/ as a power-law decay $|i-j|^{-\alpha}$ with the distance between lattice sites $i$ and $j$, where $\alpha$ is some nonnegative exponent.\footnote{The notion of {\em long-range interactions}\/ is not unanimously defined. In some communities only exponents $\alpha$ smaller than the spatial dimension of the system are called long-range. Our terminology includes these cases, but is less restrictive.} The specific long-range perturbation $\mathcal{H}_\text{pert}$ we consider is given in Eq.~\eqref{Hpert}, but other types can be treated similarly. As is well known, the transverse-field Ising chain $\mathcal{H}_\text{int}$ can be mapped onto noninteracting fermions by a Jordan-Wigner transformation, followed by a Fourier and a Bogoliubov transformation \cite{Pfeuty70,SuzukiInoueChakrabarti}, which, one might think, should bring us back onto the familiar terrain of near-integrable fermionic models. However, applying the same sequence of transformations to $\mathcal{H}_\text{pert}$ generates complicated, non-number-conserving terms beyond those that are usually considered in fermionic models. As a consequence of these additional terms, quantum kinetic equations scale less favourably with the system size and the search for an optimised truncation scheme for those equations becomes a necessity. In \cref{Selection_truncation} of this paper we introduce and discuss several such truncation schemes and benchmark them against exact results. 

The long-range part $\mathcal{H}_\text{pert}$ of the Hamiltonian can be a small perturbation for one of two reasons: either because of a small prefactor $J_z$ in \eqref{Hpert}, or because of a large value of the long-range exponent $\alpha$. The quantum kinetic theory we develop in this paper applies to both cases, but the applications and results of \cref{appli_relax} are for the latter case. To the best of our knowledge, this is the first example of a quantum kinetic perturbation theory that essentially uses $1/\alpha$ as a small parameter. The truncated set of quantum kinetic equations allows us to study the time evolution of spin expectation values and spin--spin correlation functions to high accuracy. Moreover, unlike some of the  other kinetic equations techniques, our method does not require correlation functions to factorise as in the conditions of Wick's first theorem. From our results we can distinguish different relaxation stages of the model, including prethermalisation due to the integrable part of the Hamiltonian, as well as the onset of thermalisation caused by the integrability-breaking terms.

\section{Time evolution equations of a long-range spin chain}
\label{s:dynamics}

\subsection{Near-integrable transverse-field Ising chain}
\label{Model}
We consider the Hamiltonian
\begin{equation} \label{Hamiltonien}
\mathcal{H} =\mathcal{H}_\text{int}+\mathcal{H}_\text{pert},
\end{equation}
where
\begin{equation} \label{Hint}
\mathcal{H}_\text{int} =J_x  \sum_{l}{\mathcal{S}_{l}^{x}\mathcal{S}_{l+1}^{x}} + h \sum_{l}{\mathcal{S}_{l}^{z}}
\end{equation}
describes an integrable transverse-field Ising chain, and
\begin{equation} \label{Hpert}
\mathcal{H}_\text{pert} =\frac{J_z}{2} \sum_{l,m}{\frac{1}{d(m)^{\alpha}} \mathcal{S}_{l}^{z}\mathcal{S}_{l+m}^{z}}
\end{equation}
is a long-range contribution. However, the methods developed in the following are expected to be applicable to a broader class of perturbations. Here, $l\in\llbracket 1,N\rrbracket$ labels the sites of a chain of length $N$. To each lattice site $l$ a spin-$1/2$ operator $\boldsymbol{\mathcal{S}}_l=(\mathcal{S}_l^x,\mathcal{S}_l^y,\mathcal{S}_l^z)$ is associated, satisfying the commutation relations $\boldsymbol{\mathcal{S}}_{l} \times \boldsymbol{\mathcal{S}}_{q} = i \delta_{l,q} \boldsymbol{\mathcal{S}}_{l}$ (in units of $\hbar\equiv1$). We assume periodic conditions $\boldsymbol{\mathcal{S}}_{N+1} \equiv \boldsymbol{\mathcal{S}}_1$, so that $\mathcal{H}$ is translationally invariant. Additionally, the Hamiltonian is invariant under the $\mathbb{Z}_2$ symmetry $x \to -x$. To account for the periodic boundary conditions, we define the distance between lattice sites $l$ and $l+m$ as the shortest connection around the circle, $d(m)=\min(m,N-m)$. The long-range interactions in \eqref{Hpert} decay like a power law $d(m)^{-\alpha}$ with the distance, where $\alpha$ is some nonnegative exponent. To enforce that this term contains exclusively interactions beyond nearest neighbours, the sum over $m$ extends over $\llbracket 2,N-2\rrbracket$ only. The magnetic field strength is denoted by $h$, and $J_x$ and $J_z$ are pair coupling constants.

The integrable part $\mathcal{H}_\text{int}$ of the Hamiltonian is known to be exactly solvable by a Jordan-Wigner transformation, followed by a Fourier and a Bogoliubov transformation \cite{Pfeuty70,SuzukiInoueChakrabarti} (see \cref{trans_int}). By means of this procedure, the integrable part \eqref{Hint} of the Hamiltonian can be brought into the quadratic form
\begin{equation} \label{Hint_diagonal}
\mathcal{H}_\text{int}=\sum_k{\epsilon_k\left(\eta_k^{\dag}\eta_k-\tfrac{1}{2}\right)}
\end{equation}
with dispersion relation\footnote{In case of a magnetic field reversal, such as the one we will use in \cref{particle_hole_trans}, this formula should be modified by replacing $\epsilon_k \to - \epsilon_k$ (c.f. \cref{trans_int}), which has an effect on the dynamics if $\mathcal{H}_{\mathrm{pert}} \neq 0$.}
\begin{equation} \label{dispersion_simple}
\epsilon_k = \sqrt{h^2+hJ_x\cos k+J_x^2/4},
\end{equation}
where $k$ labels the momenta in the first Brillouin zone. $\eta_k^{\phantom{\dagger}}$ and $\eta_k^\dagger$ are fermionic operators satisfying the anticommutation relations $\bigl\{\eta_k^{\phantom{\dagger}},\eta_{k'}^\dagger\bigr\}=\delta_{k,k'}$ and $\bigl\{\eta_k^{\phantom{\dagger}},\eta_{k'}^{\phantom{\dagger}}\bigr\}=0$. It is crucial for what follows to also express the perturbation $\mathcal{H}_\text{pert}$ in this preferred fermionic basis in which $\mathcal{H}_\text{int}$ is quadratic and diagonal. This guarantees that, when making approximations by neglecting high-order terms, the error will be small. A proper definition of the notion of high-order operators is given in \cref{EOM}. The main steps of transforming $\mathcal{H}_\text{pert}$ into the fermionic quasi-particle basis $\eta^{\dag}_k, \eta_k^{\phantom{\dagger}}$ are reported in \cref{pertu_transf}, leading to the normal-ordered fermionic representation
\begin{flalign*} \label{H_fermionic}
\mathcal{H}=&\mathcal{H}_{0}+\sum_k \left(A_{\mbox{\tiny I}}(k) \eta_{-k}^{\phantom{\dag}} \eta_k^{\phantom{\dag}} + A_{\mbox{\tiny II}}(k)\eta_k^{\dag} \eta_k^{\phantom{\dag}} +A_{\mbox{\tiny III}} (k) \eta_k ^{\dag} \eta_{-k} ^{\dag}\right)
\\
&+\sum_{\bm{k}} \left( B_{\mbox{\tiny I}}(\bm{k}) \eta_{-k_1}^{\phantom{\dag}}\eta_{k_2}^{\phantom{\dag}}\eta_{-k_3}^{\phantom{\dag}} \eta_{k_4}^{\phantom{\dag}} + B_{\mbox{\tiny II}}(\bm{k}) \eta_{k_1}^{\dag}\eta_{k_2}^{\phantom{\dag}}\eta_{-k_3}^{\phantom{\dag}} \eta_{k_4}^{\phantom{\dag}}+ ... + B_{\mbox{\tiny V}}(\bm{k}) \eta_{k_1}^{\dag}\eta_{-k_2}^{\dag}\eta_{k_3}^{\dag} \eta_{-k_4}^{\dag} \right) \numberthis
\end{flalign*}
of the full Hamiltonian. Here we have employed the vector notation $\bm{k}=(k_1,\dotsc,k_4)$, and $\sum_{\bm{k}}$ indicates a summation over all momenta $k_q$ ($q=1,\dotsc,4$) in the Brillouin zone. The coefficients $A_{\mbox{\tiny I}},\dotsc,B_{\mbox{\tiny V}}$ are defined in \cref{bog_formulas}, and they contain contributions from the coupling constants in the original Hamiltonian, as well as combinatorial contributions that arise from normal-ordering. Normal-ordering leads to significantly more complicated expressions here, but it will be crucial for identifying negligible terms in the approximation scheme of \cref{EOM}. $\mathcal{H}_0$ in \eqref{H_fermionic} denotes a term of degree zero in the fermionic operators, i.e., proportional to the identity. This term is irrelevant for the dynamics, but will be important for the definition of initial conditions. All terms in \eqref{H_fermionic} are momentum conserving due to the translational invariance of the spin model, and of even degree in the fermionic operators because of the $\mathbb{Z}_2$ symmetry. 

The Hamiltonian \eqref{H_fermionic} is of quartic degree in the fermionic operators $\eta$, $\eta^\dagger$. This is different from the conventional long-range Ising model in a transverse field \cite{DuttaBhattacharjee01,Halimeh_etal17,Jaschke_etal17,Mori19} where, instead of $\mathcal{H}_\text{pert}$, a perturbation $J_x \sum_{l,m} \mathcal{S}_{l}^{x}\mathcal{S}_{l+m}^{x}d(m)^{-\alpha} / 2$ with long-range couplings between the $x$-components of the spin operators is used, which leads to fermionic terms of arbitrarily high degree. The somewhat less conventional Hamiltonian \eqref{Hamiltonien}--\eqref{Hpert} we chose is a convenient model for studying approximation methods for the dynamics of the spin chain: all deviations from the exact dynamics are expected to be genuine effects of the approximations made in the time-evolution equations, as no approximations have to be made on the level of the Hamiltonian.

\subsection{Equations of motion} \label{EOM}
The time-evolution equation of an operator $\mathcal{O}$ in the Heisenberg picture is given by the von Neumann equation
\begin{equation} \label{vonNeumann}
i d_t \mathcal{O}=\left[\mathcal{O},\mathcal{H}\right],
\end{equation}
where $d_t\equiv \frac{d}{dt}$. To construct a quantum kinetic theory for the model \eqref{Hamiltonien}--\eqref{Hpert}, we require time-evolution equations of normal-ordered products of fermionic operators. For example, for $\mathcal{O}=\eta_k^\dagger \eta_k^{\phantom{\dagger}}$, a straightforward but tedious calculation yields
\begin{flalign*} \label{eom_nk}
i d_t \eta_k^\dagger \eta_k^{\phantom{\dagger}}=& -2 A_{\mbox{\tiny I}}(k) \eta_{-k}^{\phantom{\dagger}}\eta_k^{\phantom{\dagger}} - \mathrm{h.c.} - \sum_{\bm{k}} \left( B_{\mbox{\tiny I}}(\bm{k}) \Delta_{\mbox{\tiny I}}(\bm{k}) \eta_{-k_1}^{\phantom{\dagger}}\eta_{k_2}^{\phantom{\dagger}}\eta_{-k_3}^{\phantom{\dagger}} \eta_{k_4}^{\phantom{\dagger}} \right.
\\
 & + \left. B_{\mbox{\tiny II}}(\bm{k})\Delta_{\mbox{\tiny II}}(\bm{k})  \eta_{k_1}^{\dag}\eta_{k_2}^{\phantom{\dagger}}\eta_{-k_3}^{\phantom{\dagger}} \eta_{k_4}^{\phantom{\dagger}}  + ...+ B_{\mbox{\tiny V}}(\bm{k})\Delta_{\mbox{\tiny V}}(\bm{k}) \eta_{k_1}^{\dag}\eta_{-k_2}^{\dag}\eta_{k_3}^{\dag} \eta_{-k_4}^{\dag} \right), \numberthis 
\end{flalign*}
where $\Delta_{\mbox{\tiny I}}(\bm{k})=\delta_{k}^{-k_1}+\delta_{k}^{k_2}+\delta_{k}^{-k_3}+\delta_{k}^{k_4}$, $\Delta_{\mbox{\tiny II}}(\bm{k})=\Delta_{\mbox{\tiny I}}(\bm{k})-\delta_{k}^{-k_1}-\delta_{k}^{k_1}$, $\Delta_{\mbox{\tiny III}}(\bm{k})=\Delta_{\mbox{\tiny II}}(\bm{k})-\delta_{k}^{k_2}-\delta_{k}^{-k_2}$, etc. 
The right-hand side of Eq.~\eqref{vonNeumann} generally involves time-evolved operators distinct from $\mathcal{O}$. Therefore, to solve the equation of motion \eqref{eom_nk}, similar equations of motion have to be derived for the operators occurring on the right-hand side. In general, this will lead to a system of coupled differential equations whose number scales exponentially with the system size $N$. This is a problem of a complexity comparable to that of solving the von Neumann equation for the density operator in the Schr\"odinger picture, which is intractable already for moderate system sizes in most cases.

Our aim is to find a smaller differential system that is suitable for approximating the dynamics generated by the Hamiltonian \eqref{Hamiltonien}, while being numerically tractable for larger system sizes. For this purpose, it will be convenient to classify operators according to their {\em degree}\/ and their {\em $p$-particle number}.%
\begin{definition}
\label{deg_and_p}
Consider a product $\mathcal{O}=\mathcal{A}_1\cdots\mathcal{A}_q$ of fermionic operators $\mathcal{A}_i \in \{ \eta_{k_i}^{\dag},\eta_{k_i} \}$. Denote by $a \in \mathbb{N}$ the number of annihilation operators in $\mathcal{O}$, and by $c \in \mathbb{N}$ the number of creation operators. If all annihilation operators are to the right of all creation operators, the operator $\mathcal{O}$ is said to be normal-ordered. The degree of such a normal-ordered product is then defined as $\mathrm{deg}= a+c$, and we call the integer $p= \mathrm{max} (a,c)$ the {\em $p$-particle number} of $\mathcal{O}$.
\end{definition}
We define the class $\mathrm{C}_\text{deg}^p$ as the unique set of normal-ordered products of fermionic operators with a given degree and a given $p$-particle number, e.g.\
\begin{equation}\label{Odegp}
\mathrm{C}_2^1 =\bigl\{  \eta^{\dag}_k \eta_{k'}^{\phantom{\dagger}} ~| ~\forall k,k' \in \text{Br} \bigr\} \equiv \bigl\{  \eta^{\dag} \eta  \bigr\},
\end{equation}
where $\text{Br}$ denotes the Brillouin zone, and the rightmost expression is a slightly abusive shorthand notation. We furthermore define superclasses of, respectively, fixed $p$-particle number and degree,
\begin{equation}\label{OpOdeg}
\mathrm{C}^p \coloneqq \bigcup_{\text{deg} \in \llbracket p,2p \rrbracket} \mathrm{C}_\text{deg}^p, \qquad \mathrm{C}_\text{deg} \coloneqq \bigcup_{p \in \llbracket\lfloor\frac{\text{deg}+1}{2} \rfloor, \text{deg}\rrbracket} \mathrm{C}_\text{deg}^p.
\end{equation}
The union $\mathrm{F}=\bigcup_{p=0}^{N}  \mathrm{C}^p$, whose number of elements is exponentially large in the system size $N$, then spans the vector space of all fermionic operators acting on Fock space. To reduce the size of the system of coupled differential equations generated by \eqref{vonNeumann}, we introduce a {\em truncation} $\mathrm{T} \subsetneq \mathrm{F}$ as the union of, in general, several classes $\mathrm{C}_\text{deg}^p$. For example,
\begin{equation}\label{Texample}
\mathrm{T}=\mathrm{C}_0 \cup \mathrm{C}_1 \cup \mathrm{C}_2  \equiv \bigl\{\bbone,~\eta,~\eta^\dag,~ \eta^{\dag} \eta,~ \eta \eta,~\eta^\dagger\eta^\dagger \bigr\}
\end{equation}
corresponds to a truncation at the quadratic level, neglecting all terms of degree larger than two in the differential system. A first requirement on $\mathrm{T}$ to be a useful truncation is that it gives access to the observable(s) of interest. For instance, the spin component $\mathcal{S}_l^z$, expressed in terms of the fermionic operators $\eta_k$ and $\eta^\dagger_k$, is a linear combination of the identity and of some quadratic operators [see \cref{S_l^z_simplified}]. For simulating the dynamics of $\mathcal{S}_l^z$, the truncation must therefore contain at least the terms in \eqref{Texample}, even though that selection may not be sufficient to obtain a good approximation. Other choices of observables may require larger truncations. Additionally, for the sake of numerical efficiency, we want $\mathrm{T}$ to contain only the most relevant terms to describe the dynamics, at least for the time-window and observable of interest, and for the level of precision required. In that sense, kinetic theory can be seen as a perturbation theory which aims at structuring the set of all operators into a hierarchy, and then truncates that hierarchy at a chosen level.

Such truncation schemes, as is evident from the definitions of the degree and the $p$-particle number, are based on the concept of normal ordering, and at least some kind of ordering is required for a consistent classification of operators and the establishment of a hierarchy. Even if both, the Hamiltonian and the observable are given in normal-ordered form, the commutator in \eqref{vonNeumann} will usually create non-normally ordered terms in the system of coupled differential equations, which have to be normal-ordered before a truncation can be performed. For the applications considered in the present paper, the number of coupled differential equations typically scales like $N^2$ or $N^3$ with the system size $N$, and is therefore very large for system sizes of tens or even hundreds of spins. Hence, normal-ordering by hand is an arduous task. To avoid this, we have developed an algorithm, which we call the LKE (Linear Kinetic Equations) code, that takes care of the following tasks.
\begin{enumerate}
\renewcommand{\labelenumi}{(\roman{enumi})}
\item Symbolic calculation, for unspecified indices $k_1, k_2,\dotsc$, of the normal ordering of the commutators between all types of elements in $\mathrm{T}$. Technically this is equivalent to the derivation of Wick's second theorem \cite{Wick50,Kehrein}.
\item Use the results of (i) to derive, from \cref{vonNeumann}, the differential system $D$ for all operators $\mathcal{O}\in \mathrm{T}$.
\item For a given initial density operator $\rho$, define $X_0$ as the vector composed of all $\langle \mathcal{O} \rangle = \mathrm{Tr}(\rho \mathcal{O})$. Numerically solve the coupled linear differential equations $\dot{X}=DX$ with initial condition $X(0)=X_0$.
\item From $X(t)$, calculate the expectation value of the spin observable of interest, e.g.\ $\langle \mathcal{S}_l^z \rangle(t)$.
\end{enumerate}
Our LKE code is different from other kinetic equations techniques, like the one developed in \cite{Bertini_etal15,Bertini_etal16} for Hubbard-type lattice models, in that it does not require the conditions of Wick's first theorem to hold. Moreover, our approach gives direct access to correlation functions.

\subsection{Initial states}
\label{initialstates}
In principle, the LKE code described above is not restricted to specific initial states, but specific choices may simplify the problem by reducing the size of the differential system $D$. In particular, spatially homogeneous initial states, which are invariant under discrete lattice translations, are a convenient choice, because they simplify the fermionic representation of observables like $\mathcal{S}_l^z$ (see \cref{Slz_transf}). This symmetry, as well as other ones, can be used to reduce the size of the differential system of kinetic equations, an issue that is discussed in detail in \cref{symmetries}. In principle, and for convenience, one could choose a homogeneous initial state that has a simple form in the fermionic basis. More relevant for physical applications, however, are initial states that have a simple form in the spin basis, as in this case it is more likely that such a state can be prepared experimentally.

A homogeneous initial state with a particularly simple form in the spin basis is a fully $z$-polarised state $\ket{\downarrow \cdots \downarrow}$, defined such that it satisfies $\mathcal{S}_l^z \ket{\downarrow \cdots \downarrow}=-1/2\ket{\downarrow \cdots \downarrow}$ for all $l$. Since the time evolution is calculated in the $\eta$-basis, the initial state needs to be transformed into that basis as well. Fortunately, fully polarised spin states have a convenient expression in the fermionic language. For instance, one can show that the fully down $z$-polarized spin state is transformed into the Bogoliubov basis according to
\begin{equation} \label{general_ini}
\ket{\downarrow \cdots \downarrow}=\mathcal{G}_{\lfloor N/2 \rfloor} \ket{0},
\end{equation}
where $\ket{0}$ denotes the Bogoliubov vacuum and
\begin{equation} \label{def_Gn}
\mathcal{G}_n:=\frac{1}{W_n}\left(1+ \sum_{s=1}^{n} (-i)^s \sum_{0<k_1< \dotsb <k_s<\pi} \frac{v_{k_1}}{u_{k_1}} \cdots \frac{v_{k_s}}{u_{k_s}} \eta_{-k_s}^{\dag}\cdots\eta_{-k_1}^{\dag}\eta_{k_1}^{\dag}\cdots\eta_{k_s}^{\dag}  \right)
\end{equation}
for $n \in \llbracket1, \lfloor N/ 2\rfloor \rrbracket$. $u_k$ and $v_k$, defined in \cref{trans_int}, are the coefficients of the Bogoliubov transformation that diagonalises the integrable part of the Hamiltonian, and $W_n$ is a normalization constant defined by
\begin{equation}
W_n^2=1+\sum_{s=1}^{n} \sum_{0<k_1< \dotsb <k_s<\pi} \left( \frac{v_{k_1}}{u_{k_1}}\cdots\frac{v_{k_s}}{u_{k_s}}  \right)^2.
\end{equation}
We then define down \textit{truncated polarised states} as
\begin{equation}\label{truncstates}
\ket{\psi^n}=\mathcal{G}_{ n } \ket{0}
\end{equation}
with $n\in\llbracket1, \lfloor N/ 2\rfloor -1\rrbracket$. For small sizes and/or large magnetic field amplitudes $|h|$, any of these states is a good approximation of the ``proper'' polarised state \eqref{general_ini}. However, for large systems or small magnetic fields, the LKE code is expected to perform well only for initial states \eqref{truncstates} with small $n$, an effect that will become clearer in the context of the $p${\em -particle structure} introduced in \cref{hierarchyPparti} and further discussed in \cref{restriction}.

The symmetry properties of the truncated polarized states $\ket{\psi^n}$ can be used to further reduce the size of the differential system of kinetic equations. Firstly, these states belong to the even sector of the Fock space, and the time evolution under the Hamiltonian \eqref{H_fermionic} preserves this evenness. Secondly, fermions created by the operator $\mathcal{G}_n$ in \eqref{def_Gn} always come in pairs with opposite momenta $k_i$ and $-k_i$, and one can show that a differential system restricted to products of operators that take into account this pair structure is sufficient to describe not only the initial state, but also the time evolution of a truncated polarized state. Similar to the classes of operators defined in Eqs.~\eqref{Odegp} and \eqref{OpOdeg}, we denote by $\tilde{\mathrm{C}}_\text{deg}^p$ the set of products of fermionic operators with a certain degree and $p$-particle number, with the additional constraints of satisfying momentum conservation, belonging to the even sector of the Fock space, and taking into account the pair structure of $\mathcal{G}_n$. A detailed account of these symmetries and a definition of the symmetry-reduced classes $\tilde{\mathrm{C}}_\text{deg}^p$ is given in \cref{symmetries}.

Lastly, it is worth noting that the truncated polarised states \eqref{truncstates} do not satisfy the conditions of Wick's first theorem. This is an interesting observation because of the fact that, different from other quantum kinetic equations that can be found in the literature, our LKE code does not rely on the validity of Wick's theorem. For instance, for the state $\ket{\psi^1}$ one can show that
\begin{equation}
\bigl\langle\eta^{\dag}_{-k} \eta^{\dag}_k \eta_{-k'}^{\phantom{\dag}} \eta_{k'}^{\phantom{\dag}}\bigr\rangle-\bigl\langle\eta^{\dag}_{-k} \eta^{\dag}_{k}\bigr\rangle\bigl\langle\eta_{-k'}^{\phantom{\dag}} \eta_{k'}^{\phantom{\dag}}\bigr\rangle=\frac{v_k v_{k'}}{u_k u_{k'}}\left(W_1^{-4}-W_1^{-2}\right),
\end{equation}
and hence Wick's first theorem does not apply. Similar conclusions can be drawn for all truncated polarized states.

\section{Truncations and hierarchies}
\label{Selection_truncation}
Our aim is to establish a hierarchy between operators according to their relevance for the time evolution, and then truncate that hierarchy at a certain level in order to reduce the size of the differential system of kinetic equations and render it numerically more manageable. For instance, in a weakly-interacting classical kinetic theory, one would first select the \textit{ballistic} terms, for which the particles are non-interacting. If higher accuracy is needed one would include $2$-particle scattering terms, and so on. In this section we adapt this intuitive classical picture to the fermionic Hamiltonian \eqref{H_fermionic} and comment on the role of initial conditions for selecting a suitable truncation scheme. In \cref{benchmarking} we assess the quality of the approximations by benchmarking the results from different truncation schemes against exact results.

There is no rigorous theory that demands that hierarchies and truncation schemes be based on normal ordering, but on the more intuitive level one can reason as follows. Consider two fermionic states
\begin{equation}
\ket{\xi},\ket{\xi'} \in H^q\coloneqq\bigoplus_{i= 0}^q H^{(i)},
\end{equation}
i.e.\ both states reside in the sector of Fock space that corresponds to at most $q$ fermions. Then, for any normal-ordered product of fermionic operators $\mathcal{O}\in\mathrm{C}_\text{deg}$, it follows that $\bra{\xi} \mathcal{O} \ket{\xi'}=0$ if $\text{deg}>2q$, whereas such a matrix element can be nonzero if $\text{deg}\leq2q$. The same is not true without normal-ordering, i.e.\ for a non-normal-ordered product $\mathcal{O}$ of fermionic operators, the matrix element $\bra{\xi}\mathcal{O}\ket{\xi'}$ can be nonzero regardless of the degree of $\mathcal{O}$. This implies that, when disregarding non-normal-ordered operators of, say, $\text{deg}=4$, one is neglecting information not only about three and more fermions, but also about single fermions and pairs of fermions. This would contradict the intuitive, classical idea of a truncation scheme that we invoked at the beginning of this section. For a more detailed discussion of the reasoning behind normal ordering, see Chapter 4.1 of \cite{Kehrein}.

\subsection{Definitions of truncations}
\label{s:DefinitionsTruncations}

\paragraph{Truncations based on the degree of operators}
\label{hierarchy_deg}
Based on the discussion in the preceding paragraph, it is natural to base a hierarchy of fermionic operators on their degree. Correspondingly, a truncation scheme
\begin{equation}\label{deg_hierarchy}
\tilde{\mathrm{T}}_\text{deg}\coloneqq\bigcup_{j=0}^\text{deg} \tilde{\mathrm{C}}_j
\end{equation}
is defined such that it contains only normal-ordered products of fermionic operators up to a certain degree (and which additionally meet the symmetry requirements discussed in \cref{symmetries}). The cardinality of $\tilde{\mathrm{T}}_\text{deg}$, and therefore the number of variables in $\dot{X}=DX$, scales like $N^{\text{deg}/2}$ with the number of sites $N$. For instance, at the quartic level, we have
\begin{equation}\label{e:T4}
\tilde{\mathrm{T}}_{4}=\bigcup_{j=0}^{4} \tilde{\mathrm{C}}_{j} \equiv  \left\{  \bbone~;~\eta^\dag\eta^\dag,\eta^{\dag} \eta, \eta\eta~;~\eta^\dag \eta^\dag \eta^\dag \eta^\dag,\eta^\dag \eta^\dag \eta^\dag \eta,\eta^{\dag} \eta^{\dag} \eta \eta, \eta^{\dag} \eta\eta\eta,\eta \eta \eta \eta\right\},
\end{equation}
where the symbol $;$ is used to easily distinguish between the classes.

\paragraph{Truncations based on the $p$\textit{-}particle number} \label{hierarchyPparti}
Modifying the idea leading to the hierarchy \eqref{deg_hierarchy}, one can order the operators according to the integer
\begin{equation} \label{Max1}
p=1+\mathrm{max}\left\{q\in \mathbb{N}~|~\forall \ket{\xi},\ket{\xi '} \in H^q, ~\bra{\xi} \mathcal{O} \ket{\xi '}=0    \right\},
\end{equation}
which is precisely the $p$-particle number introduced in \cref{EOM}. %A system where this hierarchy applies will be said to present a {\em $p$-particle structure.}
The corresponding truncation is defined as
\begin{equation}\label{p_hierarchy}
\tilde{\mathrm{T}}^p\coloneqq\bigcup_{j=0}^p \tilde{\mathrm{C}}^j,
\end{equation}
which, for example, yields
\begin{equation}
\begin{split}
\tilde{\mathrm{T}}^{4}= \bigcup_{j=0}^{4} \tilde{\mathrm{C}}^j \equiv &\left\{  \bbone ~;~\eta^{\dag} \eta ~;~ \eta^{\dag} \eta^{\dag},\eta\eta,\eta^{\dag} \eta^{\dag} \eta \eta ~;~ \eta^{\dag} \eta^{\dag} \eta^{\dag} \eta,\eta^{\dag} \eta\eta\eta,\eta^{\dag} \eta^{\dag} \eta^{\dag} \eta \eta \eta ~;~  \right.
\\
 &~~ \left. \eta^{\dag} \eta^{\dag} \eta^{\dag} \eta^{\dag},\eta \eta \eta \eta,\eta^{\dag} \eta^{\dag} \eta^{\dag} \eta^{\dag} \eta  \eta,\eta^{\dag} \eta^{\dag} \eta \eta \eta \eta,\eta^{\dag} \eta^{\dag} \eta^{\dag} \eta^{\dag} \eta  \eta \eta \eta \right\}.
\end{split}
\end{equation}

The two truncations $\tilde{\mathrm{T}}^{p}$ and $\tilde{\mathrm{T}}_{\text{deg}}$ are equivalent for Hamiltonians which obey fermion number conservation (assuming that $\text{deg}=2 p$), but they differ in cases where, like in our fermionic Hamiltonian \eqref{H_fermionic}, terms like $\eta \eta$ or $\eta^\dag \eta^\dag$ create or destroy pairs of fermions. The cardinality of $\tilde{\mathrm{T}}^p$ scales like $N^{p}$ with the system size $N$.

\paragraph{Truncation based on degree and $p$\textit{-}particle number} \label{hierarchy_p_deg}
We can combine the ordering principles of the previous paragraphs in different ways. We introduce the truncation
\begin{equation} \label{interm_hierarchy}
\tilde{\mathrm{T}}_{\text{deg}}^{p} = \tilde{\mathrm{T}}_{\text{deg}-2} \cup \tilde{\mathrm{C}}_{\text{deg}}^{p},
\end{equation}
adding to the terms in $\tilde{\mathrm{T}}_{\text{deg}-2}$ only those of a specific degree and $p$-particle number. For example, for $\text{deg}=6$ and $p=3$ we have
\begin{equation}
\tilde{\mathrm{T}}_6^3 \equiv  \left\{  \bbone~;~\eta^\dag\eta^\dag,\eta^{\dag} \eta, \eta\eta~;~\eta^\dag \eta^\dag \eta^\dag \eta^\dag,\eta^\dag \eta^\dag \eta^\dag \eta,\eta^{\dag} \eta^{\dag} \eta \eta, \eta^{\dag} \eta\eta\eta,\eta \eta \eta \eta ~;~\eta^{\dag} \eta^{\dag} \eta^{\dag} \eta \eta \eta\right\}.
\end{equation}
The number of normal-ordered products in $\tilde{\mathrm{T}}_6^3$ scales like $N^3$ with the number $N$ of spins.

\subsection{Domain of validity of the truncations}
\label{validity}
The truncations introduced above are expected to yield good approximations of the dynamics for sufficiently short times. Longer times can be reached by tuning $\mathcal{H}$ and/or $\rho$ in a way such that expectation values of higher-degree fermionic operators are small. The main parameter for tuning the spin Hamiltonian \eqref{Hamiltonien}--\eqref{Hpert} is the long-range variable $\alpha$. The larger $\alpha$, the closer the fermionic version \eqref{H_fermionic} of the Hamiltonian is to the noninteracting integrable case. The smaller the interactions are, the longer it takes to build up correlations between fermions, and hence higher-degree fermionic operators remain close to their initial values for a longer time.

Another requirement for a truncation to yield a good approximation is that the initial state $\rho$ is uncorrelated or at most weakly correlated in the fermionic basis. Moreover, when using a truncation based on the $p$-particle hierarchy of \cref{hierarchyPparti}, the approximation works particularly well for initial states with a small fermion density. By means of the particle--hole transformation of \cref{particle_hole_trans} the validity can be extended to initial states having either a small fermion density or a small hole density,
\begin{equation} \label{few_particle_cond}
\mathrm{min} \left[ \Tr(\rho \mathcal{D}), 1-\Tr( \rho \mathcal{D}) \right] \ll 1/2,
\end{equation}
where
\begin{equation}\label{fermiondensity}
\mathcal{D}=\frac{1}{N}\sum_k\eta_k^\dag\eta_k^{\phantom{\dag}}
\end{equation}
is the (Bogoliubov) fermion density operator.
Note that, as observed in \cref{restriction}, for sufficiently small magnetic field amplitudes the fully-polarized state $\ket{\downarrow \cdots \downarrow}$ is expected to violate condition \eqref{few_particle_cond}. However, for a given system size $N$, one can choose $n$ sufficiently small such that the truncated polarized state $\ket{\psi^n}$ falls into the range of validity of the truncation. Similarly, the validity of the condition \eqref{few_particle_cond} can be enforced by increasing, at fixed $n$, the system size $N$.

\begin{figure}[t]\centering
%parametres python
%version_delta, minimal_diff_syst, True_4_particle, bench_kappa1_alpha3
%linewidth=1.2
%f=plt.figure(figsize=(7,2.7))
%plt.subplots_adjust(wspace=0.65)
%plt.rc('xtick', labelsize=9) 
%plt.rc('ytick', labelsize=9)
\begin{tikzpicture}
      \node (blabla) at (0,0){
        \includegraphics[scale=1]{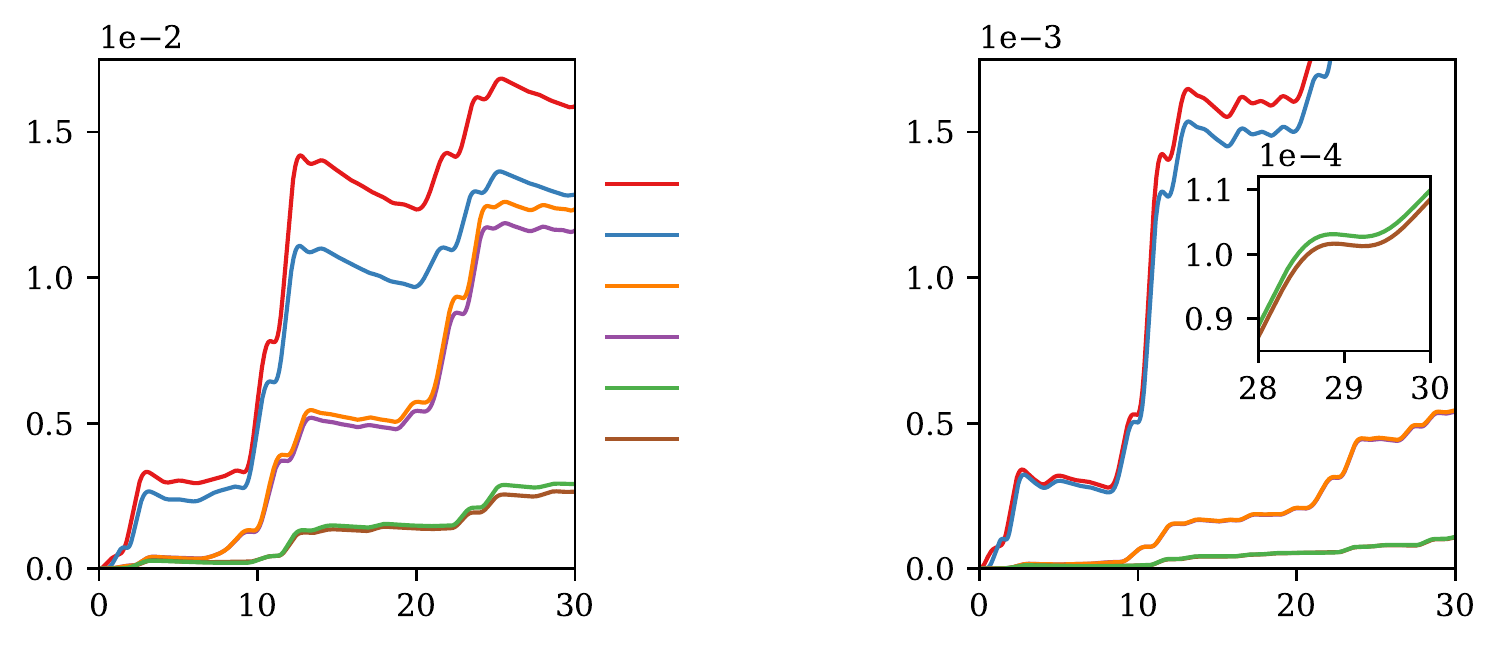}
      };
      \node[black] at (-4.12,-3.25) {$t$};
      \node[black] at (4.80,-3.25) {$t$};
      \node[black,rotate=90] at (-7.8,0.15
      7) {$\Delta\langle \mathcal{S}_l^z \rangle_{\tilde{\mathrm{T}}}$};
      \node[black,rotate=90] at (1.1,0.15
      7) {$\Delta\langle \mathcal{S}_l^z \rangle_{\tilde{\mathrm{T}}}$};
      \node[black] at (-0.3,1.45) {$\tilde{\mathrm{T}}_2$};
      \node[black] at (-0.3,0.95) {$\tilde{\mathrm{T}}^2$};
      \node[black] at (-0.3,0.4) { $\tilde{\mathrm{T}}_4$};
      \node[black] at (-0.3,-0.13) {$\tilde{\mathrm{T}}_6^{3}$};
      \node[black] at (-0.3,-0.65) {$\tilde{\mathrm{T}}^{4}_6$};
      \node[black] at (-0.3,-1.12) {$\tilde{\mathrm{T}}^4$};
\end{tikzpicture}
\caption{\label{bench_h1_final}
Comparison of the performance of the LKE code for several truncation schemes, based on the accuracy quantifier \eqref{dis_eucl} as a function of time $t$, for $\alpha=3$ (left) and $\alpha=5$ (right). Some of the truncation schemes show very similar accuracies, which indicates that irrelevant classes of fermionic operators are contained in some of them. All data are for fully-polarised initial states $\ket{\downarrow\cdots\downarrow}$ (or $\ket{\psi^{\lfloor N/2 \rfloor}}$ in the fermionic language), and for parameter values $N=10$ and $J_x=J_z=h=-1$ in the Hamiltonian.
}
\end{figure}
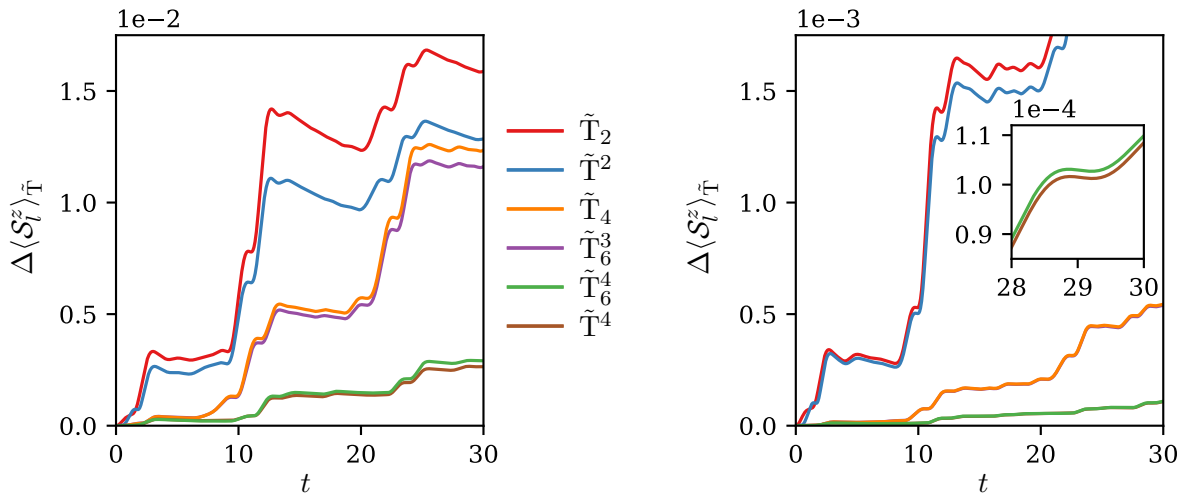

\subsection{Benchmarking}
\label{benchmarking}
In this section we assess the performance of the LKE code when using the truncations introduced in \cref{s:DefinitionsTruncations}. We compare to exact diagonalization (ED) results \cite{Weinberg2018} for system sizes up to $N=12$. As a measure for the accuracy, we use an indicator proportional to the time-integrated Euclidean distance between the LKE expectation value and the ED expectation value,
\begin{equation} \label{dis_eucl}
\Delta\langle \mathcal{S}_l^z \rangle_{\tilde{\mathrm{T}}}(t)=\sqrt{ \frac{\int_0^t \left| \langle \mathcal{S}_l^z \rangle_{\tilde{\mathrm{T}}}(u) - \langle \mathcal{S}_l^z \rangle_\text{ED}(u) \right|^2 du}{1+\int_0^t \left| \langle \mathcal{S}_l^z \rangle_\text{ED}(u) \right|^2 du}}.
\end{equation}
Based on the results for various truncation schemes shown in Fig.~\ref{bench_h1_final}, we make the following observations:
For sufficiently short times, the accuracies of the truncations follow the hierarchy
\begin{equation}
\tilde{\mathrm{T}}_2  \preceq \tilde{\mathrm{T}}^2 \prec \tilde{\mathrm{T}}_4  \preceq \tilde{\mathrm{T}}_6^3 \prec \tilde{\mathrm{T}}^4_6 \preceq \tilde{\mathrm{T}}^4,
\end{equation}
where the symbol $a \prec b$ means that the truncation $a$ is less accurate than $b$, and $\preceq $ means that the related truncations are equivalent in the large $\alpha$ limit. We note that the intuitive idea of a hierarchy based on the $p$-particle number and the degree is confirmed,\footnote{A more detailed benchmarking, which we do not show here, reveals that $\tilde{\mathrm{T}}^{p-1} \prec \tilde{\mathrm{T}}^p$ for $1 \leq p \leq 4$ on the one hand, and $\tilde{\mathrm{T}}_{\text{deg}-2} \prec \tilde{\mathrm{T}}_{\text{deg}}$ for $2 \leq \text{deg} \leq 6$ on the other hand, providing evidence of both, a degree hierarchy and a $p$-particle hierarchy. However, after the quartic level, such schemes are coarse, and it is the purpose of \cref{bench_h1_final} to propose intermediate levels of approximation.} but that ``shortcuts'' seem to exist, i.e.\ lower-order truncations that achieve more or less the same level of accuracy. For instance, $\tilde{\mathrm{T}}_6^4$ scales like $N^3$ with the system size, whereas $\tilde{\mathrm{T}}^4$ involves differential systems of size $O (N^4)$, but for the model we study these two schemes become equivalent for large $\alpha$. Similarly, $\tilde{\mathrm{T}}_4$ and $\tilde{\mathrm{T}}_6^3$ give results of essentially the same accuracy, although the first truncation contains a significantly smaller number of operators, and is therefore numerically favourable. We found these observations to hold for all system sizes $N \leq 12$ for which we had ED results available for comparison, and we do not see any reason why the observed patterns should not remain valid for larger systems with otherwise similar parameter values.

As a rule of thumb, on a regular desktop computer we can deal with system sizes $\sim 10^3$ when using a truncation scheme for which the corresponding differential system in the LKE code scales linearly with $N$; system sizes of order $\sim 10^2$ when the scaling is quadratic in $N$; and sizes of order $\sim 40$ in the case of cubic scaling. Quadratic truncations, while scaling linearly with $N$, cannot capture effects beyond integrability, and hence are not suitable for our purposes. In the following we use the compromise $\tilde{\mathrm{T}}_4$, which scales quadratically in $N$,\footnote{Another promising quartic choice that scales quadratically with $N$ is $\tilde{\mathrm{T}}_2 \cup \tilde{\mathrm{C}}_{4}^{2} \cup \tilde{\mathrm{C}}_{4}^{3} \preceq \tilde{\mathrm{T}}^3$.} as our default truncation for the applications discussed in \cref{appli_relax}.

\section{Prethermalisation and thermalisation in the long-range Ising chain} \label{appli_relax}
A nonintegrable isolated quantum system of large but finite size is expected to thermalise in a probabilistic sense, meaning that, at sufficiently late times, the expectation value $\langle \mathcal{O}\rangle(t)$ of a physically reasonable observable $\mathcal{O}$ is very close to its thermal equilibrium expectation value $\langle \mathcal{O}\rangle_{\text{th}}$ for most $t$ \cite{Goldstein_etal2010,Tasaki16}. Fluctuations around equilibrium are present, but their size is suppressed for large system sizes $N$; and while large deviations from equilibrium may occur, they are extremely rare.

In the transverse-field Ising chain with long-range interactions \eqref{Hamiltonien}--\eqref{Hpert}, the integrability of $\mathcal{H}_\text{int}$ is broken by the presence of $\mathcal{H}_\text{pert}$ which, for the large $\alpha$-values we are considering, is a weak perturbation. The relaxation to equilibrium of weakly nonintegrable systems, consisting of an integrable part plus a small nonintegrable perturbation, has been studied extensively in the literature (see \cite{Mori_etal18} and references therein). For such systems, an out-of-equilibrium initial state typically approaches equilibrium in two stages \cite{ReimannDabelow19}: On a rather short timescale, a long-lasting {\em prethermalised}\/ nonequilibrium state is reached. This state is described by a so-called {\em generalised Gibbs ensemble}\/ (GGE) \cite{Rigol_etal07,VidmarRigol16}, which, in addition to conservation of energy, takes into account also all the other conserved local charges of the integrable part of the Hamiltonian. Proper thermal equilibrium, as described by the ordinary Gibbs ensemble, is expected to be approached only much later, once the integrability-breaking perturbation becomes relevant. 

We expect a similar behaviour in the transverse-field Ising chain with long-range interactions, but with the difference that $\mathcal{H}_\text{pert}$ in \eqref{Hpert} contains integrable as well as nonintegrable contributions, as is evident from the presence of quadratic as well as quartic terms in the fermionic Hamiltonian \eqref{H_fermionic}. Both types of contributions are of small magnitude, controlled by a combination of the parameters $J_z$ and $\alpha$. Notwithstanding the similar magnitudes of the integrable and nonintegrable contributions in $\mathcal{H}_\text{pert}$, the two terms will have different effects on the equilibration of the system. The integrable portion of $\mathcal{H}_\text{pert}$ will contribute a small shift to the GGE that is reached in the initial relaxation step due to $\mathcal{H}_\text{int}$. The nonintegrable portion of $\mathcal{H}_\text{pert}$ is generically expected to effect proper thermalisation to a Gibbs state on a timescale proportional to the squared inverse of the magnitude of the nonintegrable term \cite{Mori_etal18}.

In the following we make use of the LKE code with a suitable truncation scheme in order to probe the relaxation dynamics of the transverse-field Ising chain with long-range interactions. As our local observable of interest we choose $\mathcal{S}_l^z$, the $z$-component of the spin at site $l$. This observable has the advantage of being of a simple form not only in the spin framework, but also in the fermionic language, where it is a quadratic operator \eqref{S_l^z_simplified}.

\subsection{Quadratic fermionic Hamiltonian}
\label{quad_trunc}
To distinguish effects of nonintegrability from those of the integrable model, we define the Hamiltonian
\begin{equation}\label{e:H2}
\mathcal{H}_{2}=\mathcal{H}_{0}+\sum_k \left( A_{\mbox{\tiny I}}(k) \eta_{-k}^{\phantom{\dag}} \eta_k^{\phantom{\dag}} + A_{\mbox{\tiny II}}(k)\eta_k^{\dag} \eta_k^{\phantom{\dag}} +A_{\mbox{\tiny III}} (k) \eta_k ^{\dag} \eta_{-k} ^{\dag}\right)
\end{equation}
consisting of only the quadratic terms in the fermionic Hamiltonian \eqref{H_fermionic}. $\mathcal{H}_2$ differs from $\mathcal{H}_{\text{int}}$ for finite $\alpha$, owing to the fact that $\mathcal{H}_\text{pert}$ contains not only quartic, but also quadratic contributions. For any quadratic Hamiltonian, the truncation scheme $\tilde{\mathrm{T}}_2$ yields exact results,\footnote{This is a consequence of the fact that $i d_t \mathcal{O} \in \Span\tilde{\mathrm{T}}_2$ for all $\mathcal{O} \in \tilde{\mathrm{T}}_2$; see \cref{symmetries}.} and the use of such a low-order truncation scheme will allow us to deal with fairly large system sizes of $\mathcal{H}_2$ in the following.

\begin{figure}[t]\centering
%parametres python
%version_delta, minimal_diff_syst, True_4_particle, bench_kappa1_alpha3
%linewidth=1
%f=plt.figure(figsize=(4,2))
%plt.subplots_adjust(wspace=0.65)
%plt.rc('xtick', labelsize=9) 
%plt.rc('ytick', labelsize=9)
\begin{tikzpicture}
      \node (blabla) at (0,0){
        \includegraphics[scale=1]{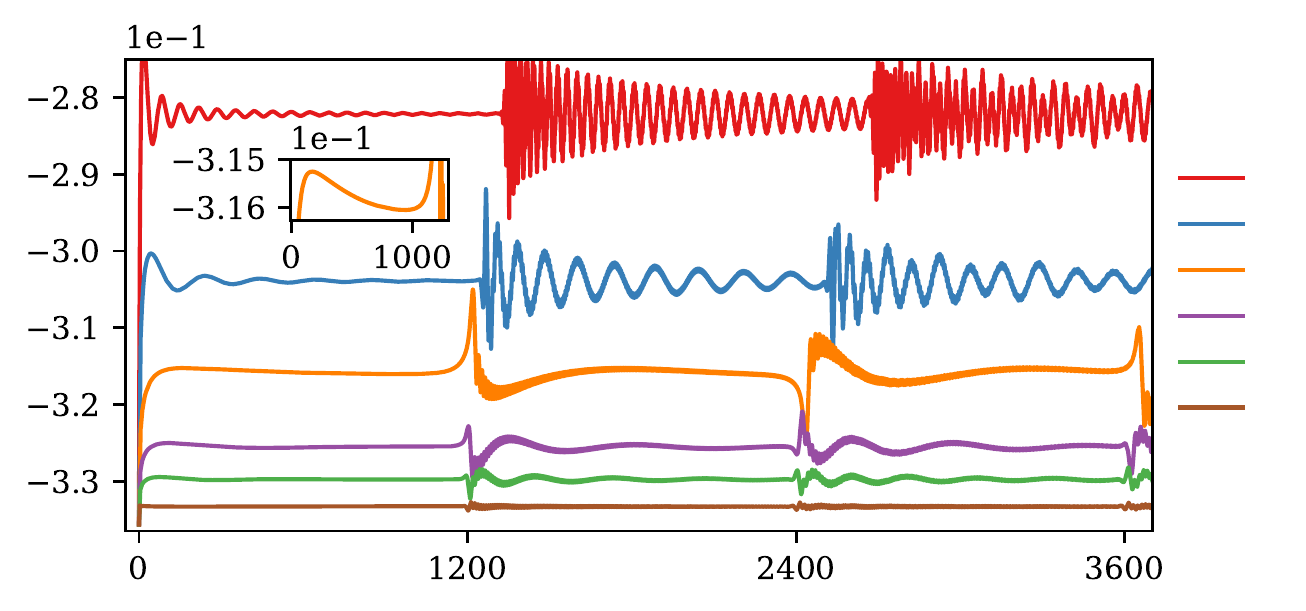}
      };
      \node[black] at (0,-3.05) {$t$};
      \node[black,rotate=90] at (-6.85,0.15
      7) {$\langle \mathcal{S}_l^z \rangle$};
      \node[black] at (6.75,1.325) {$\alpha = 4$};
      \node[black] at (6.75,0.85) {$\alpha = 5$};
      \node[black] at (6.75,0.39) { $\alpha = 6$};
      \node[black] at (6.75,-0.07) {$\alpha = 7$};
      \node[black] at (6.75,-0.55) {$\alpha = 8$};
      \node[black] at (6.85,-1.01) {$\alpha = 30$};
       
\end{tikzpicture}
\caption{Time evolution of $\langle \mathcal{S}_l^z \rangle$ under the dynamics generated by $\mathcal{H}_2$ for system size $N=1200$ with parameter values $J_x=J_z=-1$ and $h=-0.51$. The colours represent different values of the long-range parameter $\alpha$, as indicated in the legend. Time evolution starts from a truncated polarised initial state $\ket{\psi^1}$ and is calculated with the LKE code using the truncation $\tilde{\mathrm{T}}_2$, which yields exact results for the quadratic Hamiltonian considered. The main features that can be observed are a rapid initial relaxation on a timescale of the order 1, followed by a prethermalisation plateau, which can be observed until finite-size traversals obfuscate the relaxation at around $t \simeq 1200$; see main text for details.}
\label{quadratic_truncation}
\end{figure}
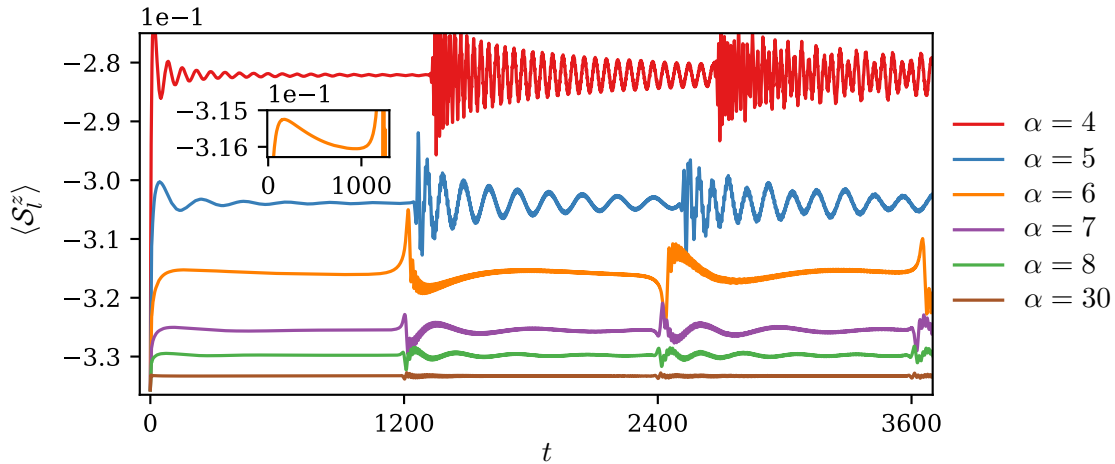

Using that scheme in the LKE code, we show in Fig.~\ref{quadratic_truncation} exact results for the time evolution of $\langle \mathcal{S}_l^z\rangle$ generated by the quadratic Hamiltonian $\mathcal{H}_{2}$ for various values of the long-range parameter $\alpha$. The most striking feature in this plot are the drastic changes, occurring periodically in the time evolution with a period of approximately 1200. These features have been termed {\em traversals}\/ in \cite{EsslerFagotti16}, and they can be understood as a finite-size effect: The dynamics is controlled by pairs of quasiparticles travelling across the chain in opposite directions, and for $\mathcal{H}_\text{int}$ the maximum velocity of quasiparticle propagation is known \cite{CalabreseEsslerFagotti12}. Because of the periodic boundary conditions, the effect of returning quasiparticles that have travelled the full length of the circle will be felt after a time $\tau_\text{trav} \simeq  N / |J_x|$ and multiples thereof. For $\alpha<\infty$ this timescale changes only slightly. As Fig.~\ref{quadratic_truncation} illustrates, the traversals spoil the relaxation behaviour that is visible up to $t\simeq 1200$. In this way, finite system sizes limit the timescales that can be assessed. For the quest of observing equilibration, which occurs on the slowest relevant timescale of a system, this poses a challenge. 

From now on we will focus exclusively on times $t$ up to $\tau_\text{trav}$. In that time window we observe in Fig.~\ref{quadratic_truncation} a rapid rise of $\langle \mathcal{S}_l^z\rangle$ from approximately $ -0.335$ [c.f.\ \eqref{Slz_psi1} and the discussion in \cref{spin_initial_mean_val}] to an $\alpha$-dependent value around $-0.3$. We estimate the corresponding timescale to be
\begin{equation} \label{time_constants}
\tau \simeq N\Big/\sum_k \left| \epsilon_k \right|,
\end{equation}
which yields a value $\tau \simeq 1$ for the parameters of Fig.~\ref{quadratic_truncation}, in agreement with the results shown in the plot. After that fast initial rise, a prethermalisation plateau is reached. We expect, but have not explicitly confirmed, that the attained long-time values agree with the GGE equilibrium values of $\mathcal{H}_2$ for the initial state used.

\subsection{Beyond integrability}
\label{weak_int_breaking}
In this section we go beyond integrability by considering the full nonquadratic fermionic Hamiltonian \eqref{H_fermionic}, which is equivalent to the long-range spin model \eqref{Hamiltonien}--\eqref{Hpert}. In this case the quadratic truncation scheme $\tilde{\mathrm{T}}_2$ is not sufficient anymore, and we opt instead for using $\tilde{\mathrm{T}}_4$ as a compromise between accuracy and numerical efficiency (see \cref{benchmarking}). $\tilde{\mathrm{T}}_4$ scales quadratically with the system size, which restricts the system sizes we can deal with on a regular desktop computer to $N=120$.  Because of the traversals discussed in \cref{quad_trunc}, this will limit the timescales that can faithfully be observed to $\tau_\text{trav}\simeq120$.

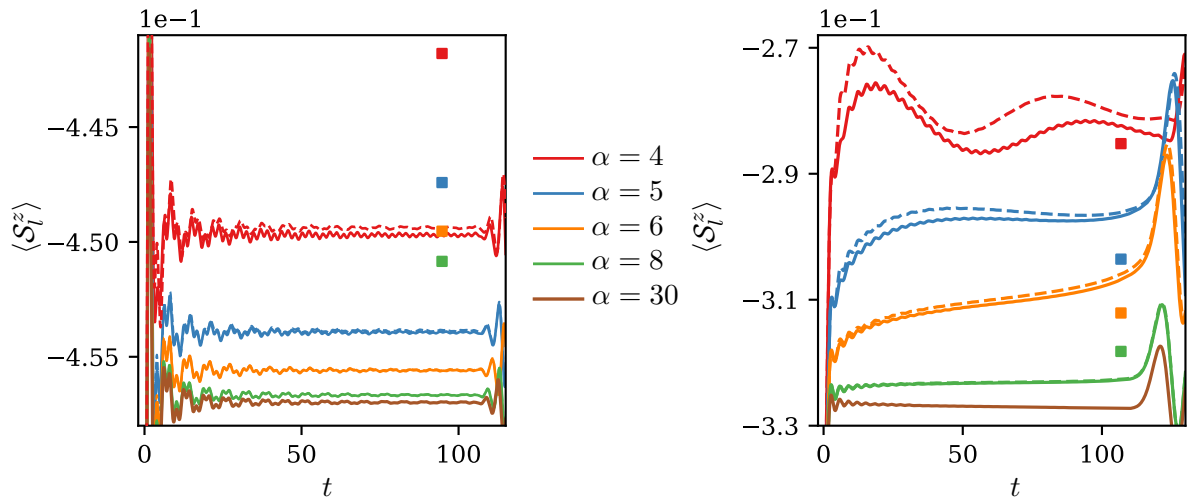
\begin{figure}[tb]\centering
%parametres python
%version_delta, minimal_diff_syst, True_4_particle, bench_kappa1_alpha3
%linewidth=1.5
%f=plt.figure(figsize=(7.7,3))
%plt.subplots_adjust(wspace=0.65)
%plt.rc('xtick', labelsize=9) 
%plt.rc('ytick', labelsize=9)
\begin{tikzpicture}
      \node (blabla) at (0,0){
        \includegraphics[scale=1]{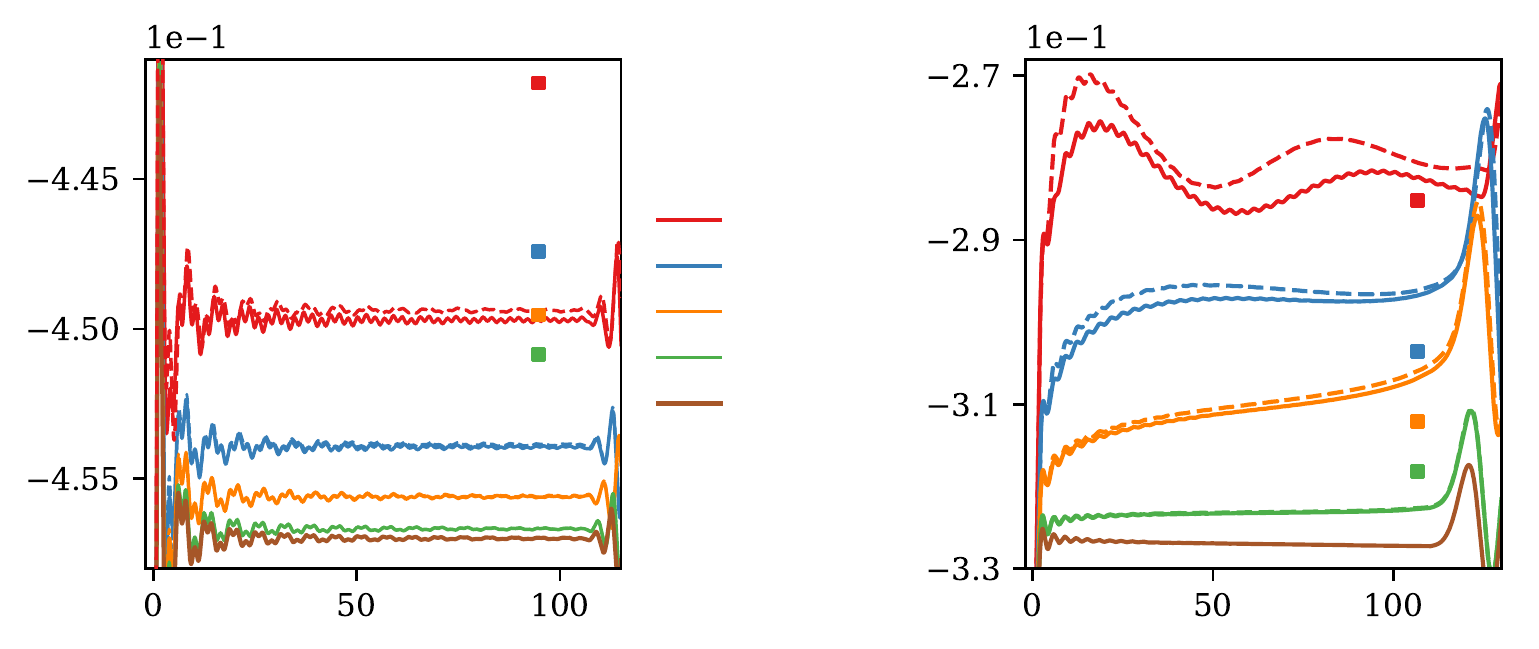}
      };
      \node[black] at (-3.85,-3.3) {$t$};
      \node[black] at (5.2,-3.3) {$t$};
      \node[black,rotate=90] at (-7.8,0.15) {$\langle \mathcal{S}_l^z \rangle$};
      \node[black,rotate=90] at (1.15,0.15) {$\langle \mathcal{S}_l^z \rangle$};
      \node[black] at (0.1,1.51-0.42) {$\alpha = 4$};
      \node[black] at (0.1,1.05-0.42) {$\alpha = 5$};
      \node[black] at (0.1,0.59-0.42) {$\alpha = 6$};
      \node[black] at (0.1,0.13-0.42) {$\alpha = 8$};
      \node[black] at (0.2,-0.33-0.42) {$\alpha = 30$};
       
\end{tikzpicture}
\caption{Time evolution of $\langle \mathcal{S}_l^z \rangle$ for system size $N=120$, coupling constants $J_x=J_z=-1$, and magnetic fields $h=-1$ (left) and $h=-0.51$ (right). Solid lines correspond to the dynamics under the full nonintegrable Hamiltonian $\mathcal{H}$ \eqref{H_fermionic}, starting from the initial state $\ket{\psi^1}$ and obtained with the LKE code and the truncation $\tilde{\mathrm{T}}_4$. The dotted lines show the exact dynamics under the quadratic Hamiltonian $\mathcal{H}_2$ using the truncation $\tilde{\mathrm{T}}_2$. Where dotted lines are not visible, they coincide with their solid counterparts. Thermal equilibrium expectation values are shown as coloured squares. As expected, a first traversal occurs around $t\simeq N$. 
} \label{CampagneN120}
\end{figure}

In Fig.~\ref{CampagneN120} the time evolution of $\langle \mathcal{S}_l^z \rangle$ is shown for various values of the long-range parameter $\alpha$, for the full nonintegrable Hamiltonian $\mathcal{H}$ as well as for the quadratic Hamiltonian $\mathcal{H}_2$. The left panel of Fig.~\ref{CampagneN120} shows that, for transverse magnetic field $h=-1$, the dynamics under $\mathcal{H}$ (solid lines) and $\mathcal{H}_2$ (dashed lines) are almost indistinguishable. Prethermal values are rapidly reached, and no subsequent drift towards the thermal values (indicated by straight solid lines) is evident on the accessible timescale. For $h=-0.51$ (right panel of Fig.~\ref{CampagneN120}), which is close to the quantum critical point of the model,\footnote{We have not studied the location of the critical point $|h_\text{c}|$ for finite $\alpha$. For a perturbation of the form $\sum_{l,m} \mathcal{S}_{l}^{x}\mathcal{S}_{l+m}^{x}d(m)^{-\alpha}$ this question has been addressed in \cite{Jaschke_etal17}. Unlike in that case, our perturbation \eqref{Hpert} couples spin components in the magnetic field direction, and for that reason we expect that the location of the critical point remains largely unaffected. For the parameter values we use, we hence expect that $h=-0.51$ is in the paramagnetic phase for all $\alpha$, which seems to be confirmed by numerical results.} dashed and solid lines clearly differ, indicating that the quartic terms in the nonintegrable Hamiltonian $\mathcal{H}$ have a sizeable effect on the dynamics, at least for the smaller $\alpha$-values considered. Moreover, the presence of nonintegrable terms appears to promote thermalisation, shifting the time-evolving expectation values closer to their thermal equilibrium value.

The thermal equilibrium values shown in Fig.~\ref{CampagneN120} are calculated according to
\begin{equation} \label{eq_ED_thermal1}
\langle \mathcal{S}_l^z \rangle_{\text{th}}= \Tr(\mathcal{S}_l^z e^{- \beta \mathcal{H}})/Z(\beta),
\end{equation}
where $Z(\beta) = \Tr( e^{- \beta \mathcal{H}})$ is the partition function. Since we are considering an isolated system where energy is conserved, the inverse temperature $\beta$ is fixed through the initial state $\ket{\psi^1}$ implicitly via
\begin{equation} \label{eq_ED_thermal2}
\lim_{N \to \infty}\bra{\psi^1}\mathcal{H} \ket{\psi^1}/N= \lim_{N \to \infty} \Tr(\mathcal{H} e^{- \beta \mathcal{H}})/N Z(\beta),
\end{equation}
at least under the idealisation of the thermodynamic limit. If $N$ is finite but sufficiently large, we expect \eqref{eq_ED_thermal2} to still be valid. Based on this assumption we use exact diagonalisation (ED) for spin chains of up to 12 sites to determine $\beta$ and $\langle \mathcal{S}_l^z \rangle_{\text{th}}$. For the energy density $\bra{\psi^1}\mathcal{H} \ket{\psi^1}/N$, an exact expression is known, see \eqref{nu_1_down}.  ED results for several small system sizes are then extrapolated to the system sizes of interest. For a magnetic field $h=-1$, we find an inverse temperature $\beta \simeq -5$, more or less independent of the value of $\alpha$. For $h=-0.51$, $\beta$ ranges from $\simeq -8.4$ to $\simeq -10.2$ for $\alpha$ between $4$ and $8$. Smaller $\alpha$ are not considered, as the validity of the approximations in our quantum kinetic theory become questionable in that case.

Negative inverse temperatures $\beta$ are known to occur in equilibrium systems with (upper and lower) bounded energy spectra if the entropy decreases as a function of energy in the high-energy region \cite{Ramsey1956}. According to \eqref{taylor1}, in our model the transition from positive to negative $\beta$ takes place at the energy $\nu_\text{th} =0$. From Eq.~\eqref{taylor1} we furthermore find  $d_\beta \nu_\text{th}(\beta)\big|_{\substack{\beta=0}} < 0$, and it is reasonable to assume that $\beta \mapsto \nu_\text{th} (\beta )$ is monotonous,\footnote{For instance, the condition $\mathrm{sp} (\mathcal{H}) \subset \mathbb{R}_-^*$ is sufficient to obtain $d_\beta \nu_{\text{th}}(\beta)<0$. Moreover, from \eqref{dispersion_relation} we know that the eigenvalues of $\mathcal{H}_{\text{int}}$ are all strictly negative as long as $h<0$, which is always the case in this section. Perturbation theory therefore proves the strict positivity of the spectrum of the full Hamiltonian \eqref{H_fermionic} for a sufficiently large (but finite) value of $\alpha$.} which is consistent with the numerical results of \cref{CampagneN120}. %{\sf [See for instance "Notes KE code 7", Fig. 6,7]}. 
According to \eqref{energy_dens_psi1_approx} the initial states we use correspond to positive energy densities, which, by virtue of the above monotonicity argument, imply negative temperatures. In \cref{high-temper} we propose a method that allows us to tune the energy density of the initial state, and hence the effective temperature, while still using truncated polarised states and staying in the regime where our quantum kinetic theory remains valid.

\subsection{Spreading of correlations}
\label{s:correlations}
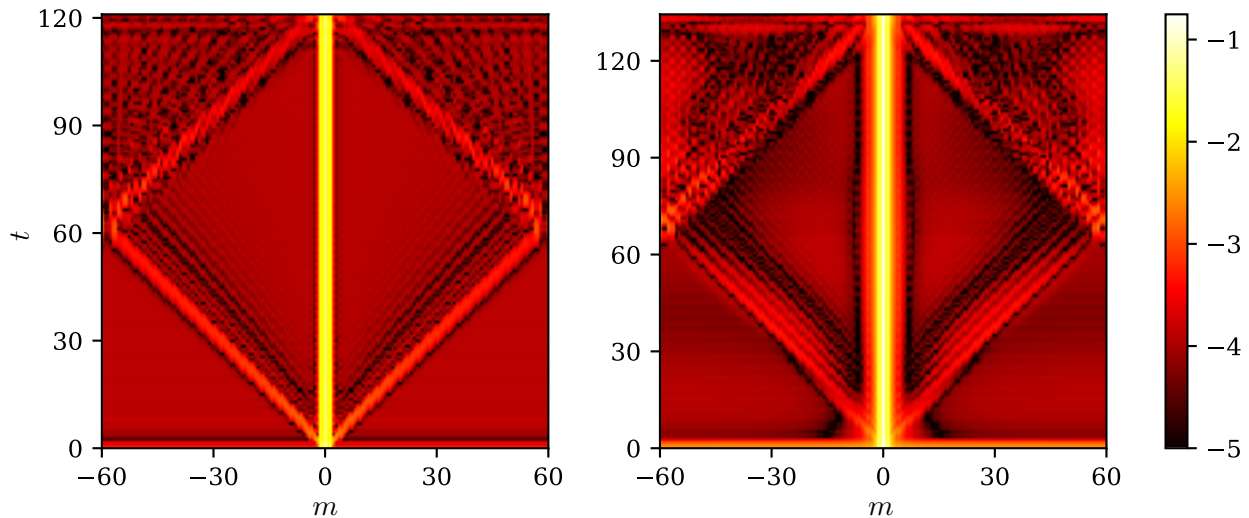
\begin{figure}[tb]\centering
%parametres python
%version_delta, minimal_diff_syst, True_4_particle, bench_kappa1_alpha3
%linewidth=1.5
%f=plt.figure(figsize=(7.7,3))
%plt.subplots_adjust(wspace=0.65)
%plt.rc('xtick', labelsize=9) 
%plt.rc('ytick', labelsize=9)
\begin{tikzpicture}
      \node (blabla) at (0,0){
        \includegraphics[scale=1]{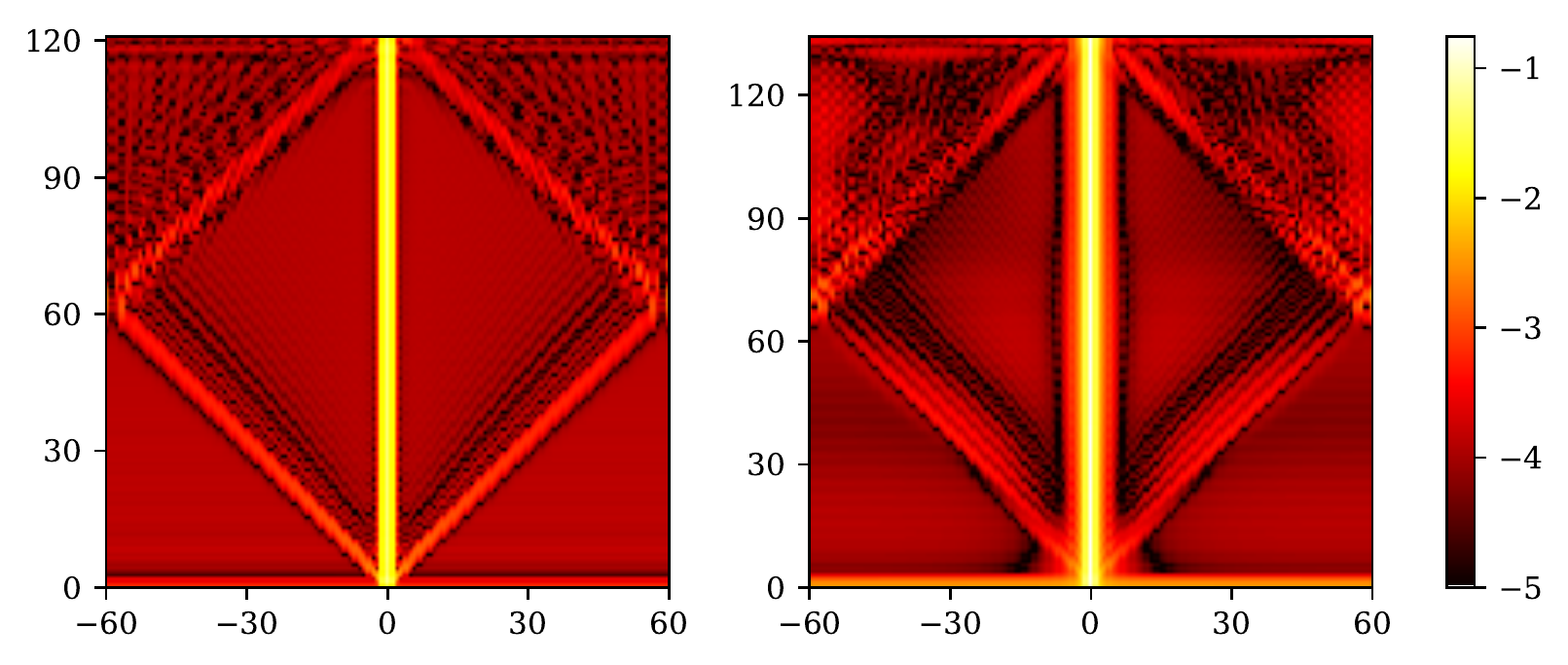}
      };
      \node[black] at (-4.13,-3.46) {$m$};
      \node[black] at (3.18,-3.46) {$m$};
      \node[black,rotate=90] at (-8.15,0.15) {$t$};
\end{tikzpicture}
\caption{Plot of $\log_{10} \left| \langle \mathcal{C}_m^z \rangle  \right|$ in the space-time plane $(m,t)$ for magnetic fields $h=-1$ (left) and $h=-0.51$ (right). The initial state is $\ket{\psi^1}$, and the time evolution is under the nonintegrable Hamiltonian \eqref{H_fermionic} with parameters $N=120$, $\alpha=4$ and coupling constants $J_x=J_z=-1$. Correlations smaller than $10^{-5}$ are irrelevant for what we want to illustrate in this plot, and we therefore rescaled the color bar to this threshold, i.e. the plots actually show $\max \left( -5, \log_{10} \left| \langle \mathcal{C}_m^z \rangle  \right| \right)$.
} \label{lightconeN120}
\end{figure}

The linear quantum kinetic theory we developed in \cref{EOM} does not make use of a Wick factorisation of correlations. This not only allows us to deal with correlated initial states, as discussed at the end of \cref{initialstates}, but also provides access to a subset of fermionic correlation functions. At the level of the $\tilde{\mathrm{T}}_4$-truncation \eqref{e:T4} that we use, we obtain all the fermionic correlations functions necessary for calculating the spin--spin correlations $\left\langle \mathcal{S}_l^x \mathcal{S}_{l+1}^x \right\rangle$, $\left\langle \mathcal{S}_l^x \mathcal{S}_{l+2}^x\right\rangle$, and $\left\langle \mathcal{S}_l^z \mathcal{S}_{l+m}^z\right\rangle$ for all $m \in \llbracket -\lfloor N/2 \rfloor,\lfloor N/2 \rfloor \rrbracket$ (see \cref{corr_function_appendix} for explicit formulas). As an example, we show in \cref{lightconeN120} the time-evolution of the connected $zz$-correlation function
\begin{equation} \label{z_correlator}
\left\langle \mathcal{C}^z_m \right\rangle=\left\langle \mathcal{S}_{l}^z \mathcal{S}_{l+m}^z \right\rangle - \left\langle \mathcal{S}_{l}^z  \right\rangle^2
\end{equation}
for $h=-1$ and $h=-0.51$, starting from the initial state $\ket{\psi^1}$. This state has nonvanishing correlations in the spin basis, which, as is visible from the short-time behaviour in  \cref{lightconeN120}, are smaller for $h=-1$, and larger for $h=-0.51$. This $h$-dependence is a consequence of the fact that tuning $h$ not only modifies the Hamiltonian \eqref{H_fermionic}, but also changes the initial state \eqref{def_Gn}, which affects the magnitude of the correlations \eqref{Slz_correlation}. As discussed in \cref{restriction}, $\ket{\psi^1}$ becomes a good approximation of $\ket{\downarrow \cdots \downarrow}$ in the limit of large negative magnetic fields, which is consistent with our observation of weaker initial correlations for a magnetic field of larger magnitude.

In \cref{lightconeN120} we show $\left\langle \mathcal{C}^z_m \right\rangle$ as a function of time $t$ and distance $m$ between lattice sites. We observe a rapid decay, on a timescale of the order one, of the nonlocal (i.e., $m$-independent) initial correlations, followed by a ``lightcone''-like spreading of correlations in space and time \cite{CalabreseCardy05,LaeuchliKollath08,Manmana_etal09,Cheneau_etal12,Bertini_etal16}. In the presence of long-range interactions, a variety of analytical, numerical, as well as experimental results indicate that, at least for sufficiently small values of $\alpha$, the linear shape of the cone gets replaced by a curved shape \cite{HastingsKoma06,EisertvdWormManmanaKastner13,Richerme_etal14,Jurcevic_etal14,FossFeigGongClarkGorshkov15,Tran_etal}. For $\alpha=4$ as used in \cref{lightconeN120} a curved shape is not visible, and it has in fact been conjectured that correlations spread strictly linearly for $\alpha$ larger than some critical value \cite{Maghrebi_etal16}. 

The spatial decay of the correlations, i.e.\ the $m$-dependence of $\left\langle \mathcal{C}_m^z \right\rangle$ at a given time $t$, appears significantly sharper in the left plot ($h=-1$) of \cref{lightconeN120} compared to the right plot ($h=-0.5$), a feature that is particularly striking at small $|m|$. This observation is consistent with the expectation that the correlation length diverges in the vicinity of the quantum critical point, which for $\alpha=4$ is expected to be close to the quantum critical point $h_\text{c}=-1/2$ of $\mathcal{H}_\text{int}$. Strictly speaking this argument is valid only for the groundstate and in equilibrium, but it is reasonable to expect signatures of quantum critically to persist at small Bogoliubov particle densities $\langle \mathcal{D} \rangle$. The initial state $\ket{\psi^1}$ has indeed a small $\langle \mathcal{D} \rangle$ \eqref{limitD} and, since the integrability breaking is weak, this remains true for fairly long times $t>0$. Furthermore, while global equilibrium has not yet been reached, regions that are some distance away from the edges of the lightcone seem to have equilibrated at least locally, with a correlation length that is presumably similar to that of the global equilibrium. Assuming all this heuristic reasoning to be valid, we interpret the qualitative differences between the two plots in \cref{lightconeN120} as consequences of the distance of the magnetic field values $h=-1$, respectively $h=-0.51$, from the quantum phase transition.

\section{Conclusions}
We have constructed quantum kinetic equations for describing the nonequilibrium dynamics of a transverse-field Ising chain with a weak integrability-breaking perturbation. The computational method we developed makes use of the Jordan-Wigner fermionic representation of the transverse-field Ising model and takes into account the integrability-breaking perturbation up to a certain degree in the time-evolution equations of operators. Which operators to include and which operators to neglect in the time-evolution equations is a crucial issue and strongly affects the accuracy of the approximation. In \cref{Selection_truncation} we have introduced, discussed, and benchmarked several truncation schemes, all of which are based on the normal-ordering of products of fermionic operators. Based on the numerical benchmarking, we found the quartic truncation scheme $\tilde{\mathrm{T}}_4$ to be numerically efficient and at the same time adequate for studying effects beyond integrability. Truncation schemes involving sixth order terms can reduce errors in time-evolved expectation values by almost an order of magnitude, but become very costly in computation time. Using the truncation scheme $\tilde{\mathrm{T}}_4$, which scales quadratically in the system size $N$, we can reach sizes of up to $N=120$ on a desktop computer, but with more effort and/or high-performance computing facilities this value can certainly be pushed quite a bit further.  

The model we have studied is the integrable transverse-field Ising chain with nearest-neighbour interactions \eqref{Hint}, with an added integrability-breaking long-range perturbation \eqref{Hpert}. The perturbation can be made small by choosing either the coupling coefficient $J_z$ in \eqref{Hpert} to be small, or the long-range exponent $\alpha$ to be large. The latter case, which we focus on in this paper, is, to the best of our knowledge, the first perturbative technique that uses $1/\alpha$ as a small parameter. 

Research on systems with long-range interactions usually focusses on one of the following two cases: (i) Systems where the long-range exponent $\alpha$ is smaller than the spatial dimension of the system. In this case a number of unconventional thermodynamic and dynamic properties are known to occur, including thermal phase transitions in one-dimensional models \cite{Dyson69a}, nonequivalence of statistical ensembles \cite{BarreMukamelRuffo01,Kastner10,KastnerJSTAT10}, and others. Our quantum kinetic theory applies to this regime when $J_z$ in \eqref{Hpert} is sufficiently small, but we have not studied this case in detail in the present work. (ii) Values of the long-range exponent $\alpha$ that are relevant for recent experiments with ultracold atoms. For spin-$1/2$ models, these are in particular $\alpha=3$ (magnetic atoms, polar molecules, Rydberg atoms) and $\alpha=6$ (Rydberg atoms); see \cite{Hazzard_etal14} for an overview. The latter case should certainly fall into the range of validity of our quantum kinetic perturbation theory when using $1/\alpha$ as a small parameter. 

An important feature of our theory is that we do not assume the conditions of Wick's first theorem to hold, i.e., unlike in some related work \cite{Bertini_etal15}, we do not reduce expectation values of quartic fermionic terms into products of expectation values of quadratic terms. This comes with some advantages and some disadvantages. A disadvantage is that we need to solve a substantially larger set of coupled differential equations, which leads to restrictions on the accessible system sizes. This is attenuated to some extend by the fact that we deal with ordinary {\em linear}\/ differential equations, whereas application of Wick's theorem leads to nonlinearities. Moreover, since quartic terms are not broken up into quadratic ones, we have access to the corresponding quantum correlation functions. In the language of spin models, this gives us access not only to spin expectation values, but also to spin--spin correlation functions, as shown in Fig.~\ref{lightconeN120}.

As an application of our perturbative scheme we studied the influence of the small parameter $1/\alpha$ on prethermalisation and thermalisation in the weakly long-range transverse-field Ising chain. Finite-size effects restrict the accessible timescales to $t \simeq N$, which in turn implies a limitation on the relaxation phenomena one can observe. Relaxation due to the integrable part $\mathcal{H}_2$ occurs on a timescale of $O(1)$ and is easily observed. Thermalisation to a Gibbs state, induced by the nonintegrable part of $\mathcal{H}_\text{pert}$, takes place on a slower timescale. While we were not able to observe the full approach to thermal equilibrium in time, we do see that the presence of nonintegrable terms pushes the spin expectation values closer to their thermal values. This effect is more pronounced closer to the quantum critical point of the model, but we do not have a satisfactory explanation for this observation.

\section*{Acknowledgement}
The authors benefited from helpful discussions with Fabian Essler, Johannes Kriel, and Stefan Kehrein. M.\,K.\ acknowledges financial support by the South African National Research Foundation through the Incentive Funding Programme and the Competitive Funding for Rated Researchers.

\appendix
\section{Transformation into the diagonal basis of \texorpdfstring{$\mathcal{H}_\text{int}$}{H int}}
As discussed in \cref{Model}, we want to express the Hamiltonian as well as observables of interest as normal-ordered products of the fermionic operators that diagonalise the integrable part of the Hamiltonian. The reasoning behind this strategy is that high-order terms in those normal-ordered operator products are expected to be less relevant for the dynamics, and the kinetic equations derived in this paper are obtained by neglecting certain classes of normal-ordered fermionic operators. In \cref{trans_int} we briefly recapitulate the standard result of diagonalising the transverse-field Ising chain with nearest-neighbour interactions by means of a Jordan-Wigner transformation, followed by a Fourier and a Bogoliubov transformation. In \cref{pertu_transf} we express the long-range contribution $\mathcal{H}_\text{pert}$ in terms of the Bogoliubov fermions of \cref{trans_int}, and in the Appendices \ref{Slz_transf} and \ref{corr_function_appendix} we do the same for spin components $\mathcal{S}_l^z$ and spin--spin correlations, which will be our observables of interest.

\subsection{Integrable part} \label{trans_int}
The integrable part $\mathcal{H}_\text{int}$ of our Hamiltonian \eqref{Hamiltonien}--\eqref{Hpert} describes a one-dimensional Ising chain in a transverse magnetic field with nearest-neighbour interactions. In this section we review the standard procedure of mapping this part of the Hamiltonian to noninteracting fermions by means of Jordan-Wigner, Fourier, and Bogoliubov transformations; see \cite{SuzukiInoueChakrabarti,CalabreseEsslerFagotti12,Sachdev2011} for more detailed accounts.

\subsubsection{Jordan-Wigner transformation}
We consider the set of operators $\mu_i^{\dag},\mu_j^{\phantom{\dag}}$ satisfying the fermionic anticommutation relations
\begin{equation}
\left\{\mu_i^{\dag},\mu_j^{\phantom{\dag}} \right\}=\delta_{i,j},\qquad\left\{\mu_i^{\phantom{\dag}},\mu_j^{\phantom{\dag}} \right\}=0=\left\{\mu_i^{\dag},\mu_j^{\dag} \right\}.
\end{equation}
Defining the Jordan-Wigner transformation
\begin{equation}
\mathcal{S}_1^+=\mu_1^{\dag},\qquad\mathcal{S}_l^+=\exp\Biggl(-i \pi \sum_{j=1}^{l-1}\mu_j^{\dag}\mu_j^{\phantom{\dag}}\Biggr)\mu_l^{\dag} \quad\text{for $l \in \llbracket 2,N\rrbracket$}
\end{equation}
with $S_l^-:=\left(S_l^+\right)^\dagger$ and $2S_l^z:=\left[S_l^+,S_l^-\right]$, it is straightforward to verify that the fermionic anticommutation relations of $\mu_i^{\dag},\mu_j^{\phantom{\dag}}$ imply that $\mathcal{S}_l^\pm,\mathcal{S}_l^z$ obey spin commutation relations, as required. Expressing the integrable part of the Hamiltonian \eqref{Hint} in terms of the fermionic operators $\mu_i^{\dag},\mu_j^{\phantom{\dag}}$ one obtains
\begin{equation}
\mathcal{H}_{\mathrm{int}}=\frac{J_x}{4} \sum_{l} \left(\mu_l^{\dag}\mu_{l+1}^{\dag} +\mu_l^{\dag}\mu_{l+1}^{\phantom{\dag}} -\mu_l^{\phantom{\dag}}\mu_{l+1}^{\dag} -\mu_l^{\phantom{\dag}}\mu_{l+1}^{\phantom{\dag}}     \right) +h \sum_{l} \left(\mu_l^{\dag} \mu_l^{\phantom{\dag}} -\tfrac{1}{2} \right).
\end{equation}

\subsubsection{Fourier transformation}
The spin Hamiltonian is invariant under discrete translations, which suggests to search for the eigenvectors of $\mathcal{H}_{\mathrm{int}}$ among the Fourier modes of a one-dimensional  lattice with periodic boundary conditions. Our convention for the discrete Fourier transformation is
\begin{equation} \label{FourierConvention}
\mu_l=\frac{1}{\sqrt{N}}\sum_k e^{ikl} \tilde\mu_k.
\end{equation}
The periodic boundaries impose conditions on the permissible momenta $k$ over which the sum in \eqref{FourierConvention} extends, dependent on the total number of fermions on the chain. This number is obtained through the operator
\begin{equation}
\mathcal{M}=\sum_l\mu_l^\dagger\mu_l^{\phantom{\dag}},
\end{equation}
and we denote its eigenvalues by $M$. Then the permissible values of the momenta are given by $k=\frac{2 \pi}{N} (q+1/2)$ with $q \in \llbracket 0, N-1  \rrbracket$ if $M$ is even, and by $k=\frac{2 \pi}{N} q$ if $M$ is odd. 

The fermionic parity (i.e., the evenness or oddness of $M$) is conserved under the time evolution not only of the integrable part, but also of the full Hamiltonian, $\left[ \mathcal{H}, e^{i \pi \mathcal{M}} \right]=0$. We will restrict our attention to initial states from the even parity sector, and parity conservation will preserve that restriction for all later times. For convenience, within that sector we shift the Brillouin zone to be as symmetric as possible around zero by choosing the integers $q \in \llbracket - \lfloor N/2 \rfloor, \lfloor N/2 \rfloor -1  \rrbracket$ in the above definition of the momenta.

To prepare for the Bogoliubov transformation to follow, we express $\mathcal{H}_\text{int}$ as a sum of matrix products,
\begin{equation}\label{Hint_muk}
\mathcal{H}_{\mathrm{int}}=-\frac{Nh}{2}+\sum_{k}{
\begin{pmatrix}
\tilde\mu_k^{\dag} & \tilde\mu_{-k}^{\phantom{\dag}}
\end{pmatrix}
R_k
\begin{pmatrix}
\tilde\mu_k \\
\tilde\mu_{-k}^{\dag}
\end{pmatrix}},
\end{equation}
where
\begin{equation}
R_k=\begin{pmatrix}
a_k & i b_k/2 \\
-i b_k/2 & 0
\end{pmatrix}
\end{equation}
with
\begin{equation}
a_k= h+\frac{J_x}{2} \cos k,\qquad b_k= \frac{J_x}{2}\sin k.
\end{equation}
Since $R_k^{\dag}=R_k^{\phantom{\dag}}$, Eq.~\eqref{Hint_muk} can be diagonalized by means of a unitary transformation.

\subsubsection{Bogoliubov transformation}
\label{Bogoliubov}
We introduce the change of basis
\begin{equation}
U_k=\begin{pmatrix}
u_k & i v_k \\
iv_k & u_k
\end{pmatrix} \in \mathrm{SU}(2)
\end{equation}
where $u_k,v_k \in \mathbb{R}$. We denote by $\eta^{\dag}_k,\eta_k $ the image of the Fourier basis under this rotation, i.e.
\begin{equation}
\begin{pmatrix}
\tilde\mu_k \\
\tilde\mu_{-k}^{\dag}
\end{pmatrix}
=
U_k^{\dag}
\begin{pmatrix}
\eta_k \\
\eta_{-k}^{\dag}
\end{pmatrix}
=
\begin{pmatrix}
u_k\eta_k -iv_k \eta_{-k}^{\dag} \\
-iv_k\eta_k +u_k \eta_{-k}^{\dag}
\end{pmatrix}.
\end{equation}
Since $\mathrm{det}(U_k)=1$, there exists a real number $x_k$ such that $u_k=\cos(x_k),~v_k=\sin(x_k)$. Requiring
\begin{equation}
U_k R_k U_k^{\dag} =\frac{a_k}{2} \bbone +\left(\frac{a_k}{2} \sin(2x_k) -\frac{b_k}{2} \cos(2x_k)   \right) \sigma^y +\left(\frac{b_k}{2} \sin(2x_k) +\frac{a_k}{2} \cos(2x_k)   \right) \sigma^z 
\end{equation}
to be diagonal yields the Bogoliubov angle $x_k=\frac{1}{2} \tan^{-1} \left(b_k/a_k \right)$. Expressing $\mathcal{H}_{\mathrm{int}}$ in terms of the thus defined fermionic operators $\eta^{\dag}_k,\eta_k $ one obtains the diagonal Hamiltonian \eqref{Hint_diagonal} with the dispersion relation
\begin{equation}\label{dispersion_relation}
\epsilon_k=\sgn(a_k)\sqrt{a_k^2+b_k^2}.
\end{equation}
This dispersion relation differs from the one in \eqref{dispersion_simple}, and also from what is given in most papers and textbooks, by the factor of $\sgn(a_k)$ \cite{Taylor1985}. This variant turns out to be useful in \cref{particle_hole_trans}, where we define a particle--hole mapping to reach the high-temperature regime, a transformation that, at least in the ferromagnetic phase, is equivalent (in the active viewpoint of symmetries) to a reversal of the magnetic field $h$.

\subsection{Perturbation} \label{pertu_transf}
In this section we express the long-range perturbation \eqref{Hpert} of the Hamiltonian in terms of the Bogoliubov fermions $\eta^{\dag}_k,\eta_k$ defined in \cref{Bogoliubov}.

\subsubsection{Jordan-Wigner and Fourier transformations}
Inserting the definitions of $\mu^{\dag}_l,\mu_l$ and $\tilde\mu^{\dag}_k,\tilde\mu_k$ into \eqref{Hpert}, we obtain
\begin{equation} \label{Hpertdec1}
\mathcal{H}_{\mathrm{pert}}=\frac{J_z\zeta_N(\alpha)}{2}  \sum_{k} \left(\tfrac{1}{4} - \tilde\mu_k^{\dag}\tilde\mu_k^{\phantom{\dag}}  \right) + \frac{J_z}{2N} \sum_{\bm{k},m}\delta_{\bm{k}}\frac{\cos[m(k_1-k_2)]}{d(m)^{\alpha}}  \tilde\mu_{1}^{\dag}\tilde\mu_{2}^{\phantom{\dag}}\tilde\mu_{3}^{\dag}\tilde\mu_{4}^{\phantom{\dag}}
\end{equation}
where
\begin{equation}
\delta_{\bm{k}}=
\begin{cases}
1 & \text{if $k_1-k_2+k_3-k_4 = 0 \mod(2\pi)$,}\\
0 & \text{else},
\end{cases}
\end{equation}
restricts the summation to momentum-conserving terms modulo $2\pi$. We have used the shorthand notation $\tilde\mu_i \coloneqq \tilde\mu_{k_i} $ and the truncated zeta-function
\begin{equation}\label{zeta_def}
\zeta_N(\alpha) \coloneqq \sum_{m=2}^{N-2} \frac{1}{d^{\alpha}}.%= \frac{3+ \left(-1 \right)^{N+1}}{2\lfloor \frac{N}{2} \rfloor ^{\alpha}}+2 \left(\zeta^{\lfloor N/2 \rfloor-1}-1 \right).
\end{equation}
%The variable $\zeta$ is defined according to $\zeta^N$, the $N$-th partial sum of the series $\sum_{k \geqslant 1} \frac{1}{k^{\alpha}}$

\subsubsection{Bogoliubov transformation} \label{bog_formulas}
Performing the Bogoliubov transformation $\tilde\mu_j=u_{j}\eta_j-iv_{j}\eta^{\dag}_{-j}$, the long-range part \eqref{Hpertdec1} of the Hamiltonian becomes
\begin{flalign*}\label{Hpert_not_ordered}
\mathcal{H}_{\mathrm{pert}}=&\frac{J_z\zeta_N(\alpha)}{2} \sum_{k} \left(\frac{1}{4}-  X_{kk} \eta_{k}^{\dag}\eta_{k}^{\phantom{\dag}} -Z'_{kk}  +i Y_{kk}(\eta_{k}^{\dag}\eta_{-k}^{\dag}- \eta_{-k}^{\phantom{\dag}}\eta_{k}^{\phantom{\dag}})  \right) \\
&+ \frac{J_z}{2N} \sum_{\bm{k},m}\delta_{\bm{k}}\frac{\cos(m(k_1-k_2))}{d^{\alpha}} \left(X_{12} \eta_{1}^{\dag}\eta_{2}^{\phantom{\dag}} +Z'_{12}\delta_{1,2}  -i Y_{12}\eta_{1}^{\dag}\eta_{-2}^{\dag} + iY_{21} \eta_{-1}^{\phantom{\dag}}\eta_{2}^{\phantom{\dag}}    \right) \\
&\qquad\qquad\qquad\qquad\qquad~~~~~\times \left(X_{34} \eta_{3}^{\dag}\eta_{4} +Z'_{34}\delta_{3,4}  -i Y_{34}\eta_{3}^{\dag}\eta_{-4}^{\dag} +iY_{43} \eta_{-3}\eta_{4}    \right)  \numberthis
\end{flalign*}
where we have defined
\begin{subequations}
\begin{align}
X_{ij} &\equiv X(k_i,k_j)=u_{k_i}u_{k_j}-v_{k_i}v_{k_j},\\
Y_{ij} &\equiv Y(k_i,k_j)= u_{k_i}v_{k_j},\\
Z_{ij} &\equiv Z(k_i,k_j)= u_{k_i}u_{k_j},\\
Z'_{ij} &\equiv Z'(k_i,k_j)= Z_{ij}-X_{ij}.
\end{align}
\end{subequations}
When normal-ordering the terms in the second sum of \eqref{Hpert_not_ordered}, further quadratic terms will emerge, which can be merged with the quadratic terms in $\mathcal{H}_{\mathrm{int}}$. The full Hamiltonian can finally be written as
\begin{flalign*} \label{H_transformed}
\mathcal{H}=&~\mathcal{H}_0+\sum_k \left( A_{\mbox{\tiny I}}(k) \eta_{-k}^{\phantom{\dag}} \eta_k^{\phantom{\dag}} + A_{\mbox{\tiny II}}(k)\eta_k^{\dag} \eta_k^{\phantom{\dag}} +A_{\mbox{\tiny III}} (k) \eta_k ^{\dag} \eta_{-k} ^{\dag} \right)
\\
&+\sum_{\bm{k}} \left( B_{\mbox{\tiny I}}(\bm{k}) \eta_{-1}^{\phantom{\dag}}\eta_{2}^{\phantom{\dag}}\eta_{-3}^{\phantom{\dag}} \eta_{4}^{\phantom{\dag}} + B_{\mbox{\tiny II}}(\bm{k}) \eta_{1}^{\dag}\eta_{2}^{\phantom{\dag}}\eta_{-3}^{\phantom{\dag}} \eta_{4}^{\phantom{\dag}}+ ... + B_{\mbox{\tiny V}}(\bm{k}) \eta_{1}^{\dag}\eta_{-2}^{\dag}\eta_{3}^{\dag} \eta_{-4}^{\dag}\right), \numberthis
\end{flalign*}
where $\mathcal{H}_0 = H_0 \,\bbone$ with
\begin{equation}\label{e:H_0}
H_0=-\frac{1}{2} \sum_k \epsilon_k+ \frac{1}{2} N J_z \zeta_N(\alpha) \Gamma_N^2  +\frac{J_z}{2N} \sum_{k,k'} c_{kk'} Y_{k'k} (Y_{kk'}+Y_{k'k})
\end{equation}
is proportional to the identity operator, and hence irrelevant for the dynamics. Here we have introduced the notations
\begin{subequations}
\begin{align}
&c_{ij} \equiv c (k_i,k_j)= \sum_{1<m<N-1} \frac{\cos[m(k_i-k_j)]}{\min(m,N-m)^{\alpha}},\\
&\Gamma_N=-\frac{1}{2}+ \frac{1}{N} \sum_{k}Z'_{kk}, \label{Gamma_N}
\\
&A_{\mbox{\tiny I}}(k) =iJ_z \zeta_N(\alpha) Y_{kk} \Gamma_N + \frac{iJ_z}{2N} \sum_{k'} c_{kk'} X_{kk'} \left(  Y_{kk'} + Y_{k'k}  \right), \label{A1def}
\\
&A_{\mbox{\tiny II}}(k) = \epsilon(k) +J_z \zeta_N(\alpha) X_{kk} \Gamma_N+\frac{J_z}{2N} \sum_{k'} c_{kk'} \left[X_{kk'}^2 -   \left( Y_{kk'}+ Y_{k'k}\right)^2 \right],
\\
&A_{\mbox{\tiny III}}(k) =-A_{\mbox{\tiny I}}(k),
\\
&B_{\mbox{\tiny I}}(\bm{k})= -\delta_{\bm{k}} \frac{J_z}{2N} c_{12} Y_{21} Y_{43},\\
& B_{\mbox{\tiny II}}(\bm{k})= 2 i\delta_{\bm{k}} \frac{J_z}{2N} c_{12} X_{12} Y_{43}, \\
&B_{\mbox{\tiny III}}(\bm{k})=2 \delta_{\bm{k}} \frac{J_z}{2N} c_{12} Y_{12} Y_{43} -\delta_{\bm{k}}\frac{J_z}{2N} c_{1,-3} \left(Z_{13}+Z'_{13}\right)\left(Z_{24} +Z'_{24}\right),\\
&B_{\mbox{\tiny IV}}(\bm{k})= -2i \delta_{\bm{k}} \frac{J_z}{2N} c_{12} X_{34} Y_{12},\\
& B_{\mbox{\tiny V}}(\bm{k})= -\delta_{\bm{k}} \frac{J_z}{2N} c_{12} Y_{12} Y_{34} . \label{B5def}
\end{align}
\end{subequations}
Note that $A_{\mbox{\tiny II}}(k) \in \mathbb{R}$ and $A_{\mbox{\tiny I}}(k),A_{\mbox{\tiny III}}(k) \in i\mathbb{R}^2$. Hermiticity imposes $\overline{A}_{\mbox{\tiny II}}=A_{\mbox{\tiny II}}$, which implies $A_{\mbox{\tiny II}}(k) \in \mathbb{R}$, but $A_{\mbox{\tiny I}}(k),A_{\mbox{\tiny III}}(k) \in i\mathbb{R}^2$ is not a necessary condition to fulfil $\overline{A}_{\mbox{\tiny I}}=A_{\mbox{\tiny III}}$. In addition, $A_{\mbox{\tiny I}}(-k)=-A_{\mbox{\tiny I}}(k)$, $A_{\mbox{\tiny III}}(-k)=-A_{\mbox{\tiny III}}(k)$ (consequence of the statistics), and $A_{\mbox{\tiny II}}(-k)=A_{\mbox{\tiny II}}(k)$. Similarly, $B_{\mbox{\tiny III}}(\bm{k}) \in \mathbb{R}$, $B_{\mbox{\tiny II}}(\bm{k}),B_{\mbox{\tiny IV}}(\bm{k}) \in i\mathbb{R}^2$, and $B_{\mbox{\tiny I}}(\bm{k}),B_{\mbox{\tiny V}}(\bm{k})\in~\mathbb{R}^2$. Hermiticity of the quartic part is guaranteed by the relations $B_{\mbox{\tiny I}}(1,2,3,4)=B_{\mbox{\tiny V}}(2,1,4,3)$, $B_{\mbox{\tiny II}}(1,2,3,4)=\overline{B}_{\mbox{\tiny IV}}(4,3,2,1)$, and $B_{\mbox{\tiny III}}(1,2,3,4)=B_{\mbox{\tiny III}}(4,3,2,1)$. Finally, $B_{\mbox{\tiny I}}(-\bm{k})=B_{\mbox{\tiny I}}(\bm{k})$, $B_{\mbox{\tiny II}}(-\bm{k})=-B_{\mbox{\tiny II}}(\bm{k})$, $B_{\mbox{\tiny III}}(-\bm{k})=B_{\mbox{\tiny III}}(\bm{k})$, with similar relations for $B_{\mbox{\tiny IV}},~B_{\mbox{\tiny V}}$.

\subsection{Transformation of \texorpdfstring{$\mathcal{S}_l^z$}{S l z}}
\label{Slz_transf}
Similarly, by performing Jordan-Wigner, Fourier, and Bogoliubov transformations, the $z$-component of the spin operator can be expressed in terms of the Bogoliubov fermions,
\begin{equation} \label{S_l^z}
\mathcal{S}_l^z=\Gamma_N \, \bbone+ \frac{1}{N} \sum_{1,2}  \left( X_{12} e^{ -i \left( k_1 -k_2  \right)l} \eta_1^{\dag} \eta_{2}
- i  Y_{21}e^{ i \left( k_1 +k_2  \right)l} \eta_{1} \eta_{2}  +i Y_{12} e^{ -i \left( k_1 +k_2  \right)l}\eta_{1}^{\dag} \eta_{2}^{\dag} \right).
\end{equation}
In general, this expressions contains also terms of the form $\langle \eta_1^{\dag} \eta_{2}\rangle$, $\langle \eta_{-1} \eta_{2}\rangle$, and $\langle \eta_{1}^{\dag} \eta_{-2}^{\dag}\rangle$, which, for $k_1\neq k_2$, are not momentum conserving. For (discrete) translationally invariant initial states, however, we show in \cref{symmetries} that such non-momentum-conserving terms have zero expectation values at all times. Therefore, the expectation value of \eqref{S_l^z} at time $t$ simplifies to
\begin{equation} \label{S_l^z_simplified}
\langle \mathcal{S}_l^z\rangle =\Gamma_N+\frac{1}{N} \sum_{k} \left(
  X_{kk} \langle \eta_k^{\dag} \eta_{k}\rangle 
+ i Y_{kk} \langle \eta_{-k} \eta_{k}\rangle    -i Y_{kk} \langle\eta_{k}^{\dag} \eta_{-k}^{\dag}   \rangle  \right).
\end{equation}
This is an important simplification for the LKE code, as it allows us to restrict the set of operators considered in the code to momentum-conserving products of Bogoliubov fermions.

\subsection{Transformation of correlation functions}
\label{corr_function_appendix}
In the spin picture we define the connected $xx$-correlation function as
\begin{equation}
\langle \mathcal{C}_1^x \rangle\equiv\langle \mathcal{S}_{l}^x \mathcal{S}_{l+1}^x \rangle - \langle \mathcal{S}_{l}^x  \rangle^2.%=\langle \mathcal{S}_{l}^x \mathcal{S}_{l+1}^x \rangle,
\end{equation}
Because of the $\mathbb{Z}_2$ symmetry of Hamiltonian and initial state we are using, we have $\langle \mathcal{S}_{l}^x  \rangle = 0$ at all times and hence $\langle \mathcal{C}_1^x \rangle=\langle \mathcal{S}_{l}^x \mathcal{S}_{l+1}^x \rangle$. By performing Jordan-Wigner, Fourier, and Bogoliubov transformations and assuming a translationally invariant state, this correlation function turns out to be quadratic when expressed in terms of the Bogoliubov fermions,
\begin{equation}\label{Slx_correlation}
\langle \mathcal{C}^x_1 \rangle =\frac{1}{2N} \sum_k \mathrm{Tr} \left[ \bm{p}(k)^T \left(\bm{q}_{0}(k)+i \bm{q}_{\mbox{\tiny I}}(k) \langle\eta_{-k}^{\phantom{\dag}} \eta_k^{\phantom{\dag}}\rangle + \bm{q}_{\mbox{\tiny II}}(k)\langle\eta_k^{\dag} \eta_k^{\phantom{\dag}}\rangle -i\bm{q}_{\mbox{\tiny I}} (k)\langle \eta_k ^{\dag} \eta_{-k} ^{\dag}\rangle \right) \right],
\end{equation}
where
\begin{equation}
\bm{p}(k)=  \begin{pmatrix} \cos(k) \\ \sin(k) \end{pmatrix},\quad \bm{q}_{0}(k)=\begin{pmatrix} Z'_{kk} \\ -Y_{kk} \end{pmatrix},\quad\bm{q}_{\mbox{\tiny I}}(k) =\begin{pmatrix} Y_{kk} \\ -X_{kk}/2 \end{pmatrix},\quad \bm{q}_{\mbox{\tiny II}}(k) =\begin{pmatrix} X_{kk} \\ 2Y_{kk} \end{pmatrix}.
\end{equation}
Similarly, correlation functions $\langle \mathcal{C}^x_m\rangle=\langle \mathcal{S}_{l}^x \mathcal{S}_{l+m}^x \rangle$ can be expressed in terms of Bogoliubov fermions for any $m$, and they turn out to be of degree $2m$ in the fermionic basis. With the $\tilde{\mathrm{T}}_4$ truncation \eqref{e:T4} we are using in Sections \ref{weak_int_breaking} and \ref{s:correlations} we have access to all quartic fermionic operators, but not to all sixth-order terms, so that we can calculate $\langle \mathcal{C}^x_2\rangle$, but not the $xx$ correlations of spins that are further than two lattice sites apart.

For the $zz$-correlation function with respect to a translationally invariant state we find
\begin{flalign*} \label{Slz_correlation}
\langle \mathcal{S}_{l}^z \mathcal{S}_{l+m}^z \rangle =&~C_0(m)+\sum_k \left( R_{\mbox{\tiny I}}(k,m) \langle\eta_{-k}^{\phantom{\dag}} \eta_k^{\phantom{\dag}}\rangle + R_{\mbox{\tiny II}}(k,m)\langle\eta_k^{\dag} \eta_k^{\phantom{\dag}}\rangle +R_{\mbox{\tiny III}} (k,m)\langle \eta_k ^{\dag} \eta_{-k} ^{\dag}\rangle \right)
\\
&+\sum_{\bm{k}} \left( S_{\mbox{\tiny I}}(\bm{k},m)\langle \eta_{-1}^{\phantom{\dag}}\eta_{2}^{\phantom{\dag}}\eta_{-3}^{\phantom{\dag}} \eta_{4}^{\phantom{\dag}}\rangle + S_{\mbox{\tiny II}}(\bm{k},m)\langle \eta_{1}^{\dag}\eta_{2}^{\phantom{\dag}}\eta_{-3}^{\phantom{\dag}} \eta_{4}^{\phantom{\dag}}\rangle+ ... + S_{\mbox{\tiny V}}(\bm{k},m)\langle\eta_{1}^{\dag}\eta_{-2}^{\dag}\eta_{3}^{\dag} \eta_{-4}^{\dag}\rangle\right) \numberthis
\end{flalign*}
with
\begin{subequations}
\begin{align}
&\gamma_{ij}(m) \equiv \gamma (k_i,k_j,m)= \cos[m(k_i-k_j)],
\\
&C_0(m) =\frac{2}{J_z N} \bigg[ \frac{1}{2} \sum_k \epsilon_k + H_0 \left( k, \left\{c_{i,j} \to \gamma_{i,j}(m), \zeta_N(\alpha) \to 1 \right\} \right) \bigg],
\\
&R_{\mbox{\tiny I}}(k,m) =\frac{2}{J_z N}  A_{\mbox{\tiny I}}\left( k,  \left\{c_{i,j} \to \gamma_{i,j}(m), \zeta_N(\alpha) \to 1 \right\} \right),
\\
&R_{\mbox{\tiny II}}(k,m) =\frac{2}{J_z N} \left[A_{\mbox{\tiny II}} \left( k, \left\{c_{i,j} \to \gamma_{i,j}(m), \zeta_N(\alpha) \to 1 \right\} \right) - \epsilon_k \right],
\\
&R_{\mbox{\tiny III}}(k,m) =-R_{\mbox{\tiny I}}(k,m),
\\
& S_{\lambda}(\bm{k},m) = \frac{2}{J_z N} B_{\lambda}(\bm{k}, \left\{c_{i,j} \to \gamma_{i,j}(m)\right\}), \qquad \text{for} ~\lambda = \mbox{\tiny I}, \dotsc, \mbox{\tiny V}.
\end{align}
\end{subequations}
Making use of $\langle \mathcal{S}_{l}^z \rangle$ \eqref{S_l^z_simplified} then allows us to express the connected $zz$-correlation function \eqref{z_correlator} in the fermionic basis. Since the resulting expression is quartic in the fermionic operators, it follows that the $\tilde{\mathrm{T}}_4$ truncation \eqref{e:T4} is sufficient for calculating $\langle \mathcal{C}^z_m\rangle$ for arbitrary $m$.

\section{Symmetries}
\label{symmetries}

In \cref{EOM} we have introduced truncations of the equations of motion based on the notions of the degree and the $p$-particle number of products of fermionic operators. The idea behind such a truncation is the knowledge (or belief, or hope) that the neglected classes of operators do not contribute significantly to the dynamics of the quantities of interest, at least on a certain time scale. Neglecting these operators drastically reduces the size of the differential system that needs to be considered. In addition to such approximately vanishing quantities, symmetries of the Hamiltonian and/or the initial state may lead to a strict decoupling of equations of motion, in the sense that the differential equations for a certain set of fermionic products is strictly independent of some other set of fermionic products (although the converse may in general not be true). Such a decoupling can, on top of the truncation that is essentially introduced by hand, further reduce the size of the differential system.

For instance, in the Schr\"odinger picture, if $\mathcal{O}$ is a normal-ordered product that does not commute with a given symmetry $\mathcal{U}$, i.e. $\mathcal{O} \in \mathrm{C}_{\cancel{\mathcal{U}}}$, and if the operators $\mathcal{U},\rho,\mathcal{H}$ verify the conditions
\begin{enumerate}
\renewcommand{\labelenumi}{(\alph{enumi})}
\item at time $t$, if $\left[\rho(t), \mathcal{U}\right] =0$, then $\Tr \left[ \rho(t) \mathcal{O} \right]=0$ holds true, \label{hyp1}
\item $[\mathcal{H},\mathcal{U}]=0$, \label{hyp2}
\item $[\rho(0),\mathcal{U}]=0$, \label{hyp3}
\end{enumerate}
then it follows\footnote{For infinitesimal $dt$, hypotheses (a) and (b) imply that $\rho(t+dt)= -i \left[ \mathcal{H}, \rho \right] dt + \rho(t)$ commutes with $\mathcal{U}$, so that $\mathrm{Tr} \left( \rho(t') \mathcal{O} \right)=0$ for all $t'>t$.} that $\mathrm{Tr} \left( \rho(t) \mathcal{O} \right)=0$ for all $t$. As a result, operators in the complement $\mathrm{C}_{\cancel{\mathcal{U}}}$ can be safely ignored, without any approximation, when constructing the differential system of kinetic equations.

As pointed out in \cref{initialstates}, the truncated polarised states \eqref{truncstates} that we use as initial states have the following symmetries and resulting conservation laws:
\begin{enumerate}
\renewcommand{\labelenumi}{(\roman{enumi})}
\item Discrete translation invariance of the initial states $\ket{\psi^n}$ and the Hamiltonian $\mathcal{H}$, resulting in conservation of %the expectation value of 
the total lattice momentum. %operator $\mathcal{P} \coloneqq \sum_k k \, \eta_k^\dag \eta_k^{\phantom{\dag}}$. 
For the example of the class of products of fermionic operators of degree 2 and $p$-particle number 1 in \eqref{Odegp}, the symmetry-reduced set is then given by $\bigl\{  \eta^{\dag}_k \eta_{k}^{\phantom{\dagger}} ~| ~\forall k \in \text{Br} \bigr\}$, which scales with system size as $N$ instead of $N^2$. Similarly, momentum conservation reduces the number of operators in any of the classes $\mathrm{C}_4^p$ to scale with system size like $N^3$ instead of $N^4$.
\item Spin inversion symmetry in the $x$-direction of the spin Hamiltonian \eqref{Hamiltonien}, leading to conservation of the fermionic parity $\langle \exp\bigl(i\pi\sum_k \eta^\dag_k \eta^{\phantom{\dag}}_k\bigr) \rangle$ in the fermionic picture. Since the truncated polarized initial states $\ket{\psi^n}$ lie entirely in the even parity sector of the Fock space, all odd-degree normal-ordered products of fermionic operators strictly do not contribute and can be excluded from the differential system of kinetic equations.\footnote{ Many odd-degree normal-ordered products are also non-momentum-conserving, and have already been eliminated in (i), e.g.\ all $\eta_k$ for which $k \not\in \left\{0,\pi \right\}$. More generally, the set $S$ of operators of degree $2j+1\leq 2N$ that can be written as normal-ordered products of a momentum-conserving product of degree $2j$ times $\eta_\pi$ has also been eliminated before, and $S$ scales like $2j-1$.}
\item Fermions created by the operator $\mathcal{G}_n$ in \eqref{def_Gn} always come in pairs with opposite momenta $k_i$ and $-k_i$. Through the definition of the truncated polarized states $\ket{\psi^n}$ in Eq.~\eqref{truncstates}, this implies that
\begin{equation}
\bra{\psi^n}\eta_k^\dag\eta_k^{\phantom{\dag}}\ket{\psi^n}=\bra{\psi^n}\eta_{-k}^\dag\eta_{-k}^{\phantom{\dag}}\ket{\psi^n}\qquad\forall k,n.
\end{equation}
\end{enumerate}
To exploit in the kinetic equations the symmetries (i)--(iii), we define, based on the classes $\mathrm{C}_\text{deg}^p$ of normal-ordered products of fermionic operators defined in \cref{EOM}, the symmetry-restricted classes
\begin{equation}\label{Ctilde}
\tilde{\mathrm{C}}_\text{deg}^p=
\begin{cases}
\bbone & \text{for $\text{deg}=0$},\\
\left\{\mathcal{P}_1\cdots\mathcal{P}_{\text{deg}/2} \right\}\in\mathrm{C}_\text{deg}^p \text{ with } \mathcal{P}_i\in\left\{\eta_{k_i}\eta_{-k_i},\eta_{k_i}^\dag\eta_{-k_i}^\dag,\eta_{k_i}^\dag\eta_{k_i}\right\} & \text{for $\text{deg},p\in2\mathbb{N}$},\\
\emptyset & \text{else}.
\end{cases}
\end{equation}
This definition excludes, in agreement with (ii), all normal-ordered products of odd degree, and the construction via pair operators $\mathcal{P}_i$ in the second line of \eqref{Ctilde} guarantees momentum conservation (i) as well as the pair structure (iii). For example, the class $\tilde{\mathrm{C}}_4^2$ contains only elements of the form $\eta^\dag_{-k} \eta^\dag_{k} \eta^{\phantom{\dag}}_{-k'} \eta^{\phantom{\dag}}_{k'}$ and $\eta^\dag_{k} \eta^\dag_{k'} \eta^{\phantom{\dag}}_{k} \eta^{\phantom{\dag}}_{k'}$. We define $\tilde{\mathrm{C}}^p$ and $\tilde{\mathrm{C}}_\text{deg}$ analogous to \eqref{OpOdeg}, and the union
\begin{equation}
\tilde{\mathrm{F}}=  \bigcup_{p=0}^{N} \tilde{\mathrm{C}}^{p}
\end{equation}
contains all normal-ordered products of fermionic operators that satisfy the symmetry restrictions (i)--(iii). A proper subset $\tilde{\mathrm{T}} \subsetneq \tilde{\mathrm{F}}$ can then serve as a truncation of the differential system of kinetic equations.

For a truncated polarized initial state $\rho= \ket{\psi^n} \bra{\psi^n}$, we have $\mathrm{Tr} \left(\rho \mathcal{O} \right)=0$ for all $\mathcal{O}\in \mathrm{F} \setminus \tilde{\mathrm{F}}$. Moreover, for all $\tilde{\mathcal{O}} \in \tilde{\mathrm{F}}$, one finds $i d_t \tilde{\mathcal{O}} \in \Span\tilde{\mathrm{F}}$.\footnote{This follows from a similar property of the $(p\leq2)$-subset $\tilde{\mathrm{T}}_2=\tilde{\mathrm{C}}_0 \cup \tilde{\mathrm{C}}_2$, where it turns out that $\left[ \mathcal{O}, \mathcal{Q}  \right] \in \Span\tilde{\mathrm{T}}_2$ for all $\mathcal{O}, \mathcal{Q} \in \tilde{\mathrm{T}}_2$.} Hence, a differential system that contains only products of operators from the set $\tilde{\mathrm{F}}$ is sufficient to describe not only the initial state, but also the time evolution of a truncated polarized state. The symmetry restrictions reduce the number of operators to be considered in the LKE code significantly, for example from $N^4$ to $N^2$ when going from $\mathrm{C}_4^p$ to $\tilde{\mathrm{C}}_4^p$.

\section{Dynamics and thermodynamics in the high-temperature limit} \label{high_temp_dynamics}

In \cref{weak_int_breaking} we employed exact diagonalisation of small, finite systems for calculating thermal expectation values, and extrapolated these values to large system sizes $N$. In this section we discuss how, without resorting to exact diagonalisation nor to conventional diagrammatic perturbation theory at equilibrium\cite{Bertini_etal16}, thermal expectation values in the high-temperature limit can be obtained for large system sizes. We calculate, up to second order in the inverse temperature $\beta$, analytical expressions for the thermal values of both the energy density $\langle \mathcal{H} \rangle/N$ and the spin observable $\langle S_l^z \rangle$.

The truncated down-polarised states $\ket{\psi^n}$ that were introduced in \eqref{truncstates}, and for which our LKE code was tailored, do not usually fall into the regime where $\beta\ll1$, and the same holds true for their up-polarized counterparts, which we denote by $\ket{\chi^n}$. However, by considering a suitable superposition of $\ket{\psi^n}$ and $\ket{\chi^n}$ the energy density can be tuned into the high-temperature regime. Under suitable conditions one can then argue that the dynamics of $\ket{\psi^n}$ and $\ket{\chi^n}$ decouples. Making use of a particle--hole transformation, the LKE code can be used for the calculation of the decoupled time evolution of $\ket{\chi^n}$, and the outcome can be compared to the thermal values obtained from a high-temperature expansion.

\subsection{Thermal equilibrium results in the high-temperature limit} \label{high-temper}
For sufficiently high temperatures, the Boltzmann factor can be expanded up to second order in $\beta$,
\begin{equation}
e^{-\beta \mathcal{H}}= 1 - \beta \mathcal{H} +\beta ^2 \mathcal{H}^2 /2 + O(\beta^3).
\end{equation}
Based on this expansion, and making use of $\Tr\left( \mathcal{H} \right)=0$, one can derive, up to the same order in $\beta$, the thermal expectation value of the energy density \cite{Lepoutre2018},
\begin{equation} \label{taylor1}
\upsilon_{\mathrm{th}}(\beta):= \Tr\left( \mathcal{H}e^{-\beta \mathcal{H}}/Z(\beta) \right) =- K_1 \beta  - K_2 \beta^2 + O(\beta^3),
\end{equation}
where $K_1=2^{-N}\Tr\mathcal{H}^2 /N > 0$ and $K_2=-2^{-N-2} \Tr\mathcal{H}^3/N$. We find
\begin{equation} \label{taylor2}
K_1= h^2 /2^2 + J_x^2/2^4 + J_z^2 \zeta_N(2 \alpha) / 2^5,
\end{equation}
with the partial zeta-function $\zeta_N$ as defined in \eqref{zeta_def}, which, for $\alpha>1/2$, converges in the large-$N$ limit. At next order, after a long calculation we find
\begin{equation} \label{blabla}
K_2=-\frac{3}{2^6} h^2 J_z \zeta_{N}(\alpha)-\frac{J_z^3}{2^8} \sum_{1<m<N-1} \frac{1}{d(m)^{\alpha}} \sum_{\substack{1<n<N-1\\  n \not\in L}} \frac{1}{d(n)^{\alpha}} \frac{1}{d(q)^{\alpha}}
\end{equation}
with $q \equiv m+n\bmod N$ and $L=\left\{ N-1-m,N-m,N+1-m \right\}$. Equations \eqref{taylor2} and \eqref{blabla} can be evaluated without too much effort for very large system sizes. For the purpose of benchmarking the LKE code, we are interested in $\alpha$ substantially larger than $1/2$. For such values we observe that $K_1$ and $K_2$ converge quickly with increasing $N$, being essentially indistinguishable from their infinite-system limit already for system sizes $\sim 10^2$. In that same regime of $\alpha$-values we also find that $K_2 \ll K_1$. In fact, we expect more generally that odd orders in $\beta$ dominate the Taylor expansion \eqref{taylor1}, because $\nu_{\mathrm{th}} (\beta)$ is approximately an odd function. Along similar lines, we obtain
\begin{equation}
\langle S_l^z \rangle_\text{th} = - \frac{h}{2^2} \beta + \frac{h J_z \zeta_{N}(\alpha)}{2^5} \beta^2 +O(\beta^3)
\end{equation}
for the observable of interest, and again the quadratic term in $\beta$ is negligible if $\alpha$ is appreciably larger than $1/2$. Based on these results, we can obtain accurate thermal expectation values $\upsilon_{\mathrm{th}}$ and $\langle S_l^z \rangle_\text{th}$ for the system sizes of order $\sim 10^2$ we want to benchmark against (or for much larger ones), in the regime of not-too-small $\alpha$ and small inverse temperatures $\beta$.

\subsection{Particle--hole conjugation}
\label{particle_hole_trans}

The truncated polarized initial states $\ket{\psi^n}$, being highly ordered, are typically outside the high-temperature regime. Our strategy for obtaining a high-temperature initial state is to define, in the fermionic picture, a unitary particle--hole transformation, which maps fermions onto empty sites and vice versa. In the spin picture, this corresponds to transforming a fully down-polarized state $\ket{\downarrow\cdots\downarrow}$ into a fully up-polarized state $\ket{\uparrow\cdots\uparrow}$. Similarly, a truncated down-polarized state $\ket{\psi^n}$ is transformed into a truncated up-polarized state $\ket{\chi^n}$, and we find that a suitably chosen linear combination of the two states falls into the high-temperature regime, as demonstrated in \cref{energy_densit} for $n=1$ and $n=\lfloor N/2 \rfloor$.

We define the particle--hole transformation as the unitary operator $\mathcal{U}$ that transforms Jordan--Wigner fermionic operators according to
\begin{equation}\label{unitary_part_hole}
\mathcal{U} \mu_l \mathcal{U}^{\dag}=\mu_l^\dag.
\end{equation}
Under this transformation, all Jordan-Wigner fermionic states are mapped onto their particle--hole counterparts, e.g.\ $\mathcal{U}\ket{1011}=\ket{0100}$. From Eq.~\eqref{unitary_part_hole} it follows that $\mathcal{U}$ acts on Bogoliubov fermionic operators as
\begin{equation}
\mathcal{U} \eta_k \mathcal{U}^{\dag}=\eta_{-k}^\dag.
\end{equation}
By means of this transformation, we define truncated up-polarized states as
\begin{equation}
\ket{\chi^n} := \mathcal{U} \ket{\psi^n}.
\end{equation}
To understand how the Hamiltonian is transformed under $\mathcal{U}$, we note that, in the spin framework, the spin operators are transformed like
\begin{equation}
(\mathcal{S}_l^x,\mathcal{S}_l^y,\mathcal{S}_l^z) \mapsto ((-1)^{l+1}\mathcal{S}_l^x,(-1)^{l+1}\mathcal{S}_l^y,-\mathcal{S}_l^z).
\end{equation}
Transforming the Hamiltonian \eqref{Hamiltonien} under $\mathcal{U}$ corresponds to a sign reversal of constants,
\begin{equation}
\mathcal{U} \mathcal{H}(h,J_x)\mathcal{U}^\dag=\mathcal{H}(-h,-J_x).
\end{equation}
In the case of $2|h|>|J_x|$, to which we apply the LKE code in this paper, we have $\sgn(2h+J_x\cos k)=\sgn (h)$ for all momenta $k$, which implies that, according to \eqref{dispersion_relation}, the magnetic field reversal effected by $\mathcal{U}$ amounts to replacing $\epsilon_k \to - \epsilon_k$ in $\mathcal{H}_{\mathrm{int}}$. For the perturbation $\mathcal{H}_{\mathrm{pert}}$, the dictionary \eqref{A1def}--\eqref{B5def} remains unchanged.

\subsection{Decoupled dynamics}
\label{decoupled_dynamics}
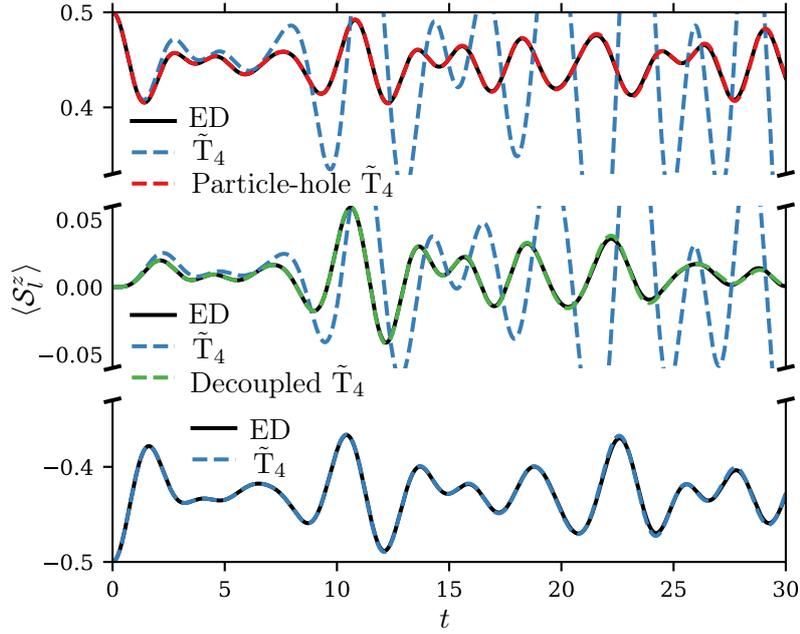
\begin{figure}[tb]\centering
\begin{tikzpicture}
      \node (blabla) at (0,0){
        \includegraphics[scale=1]{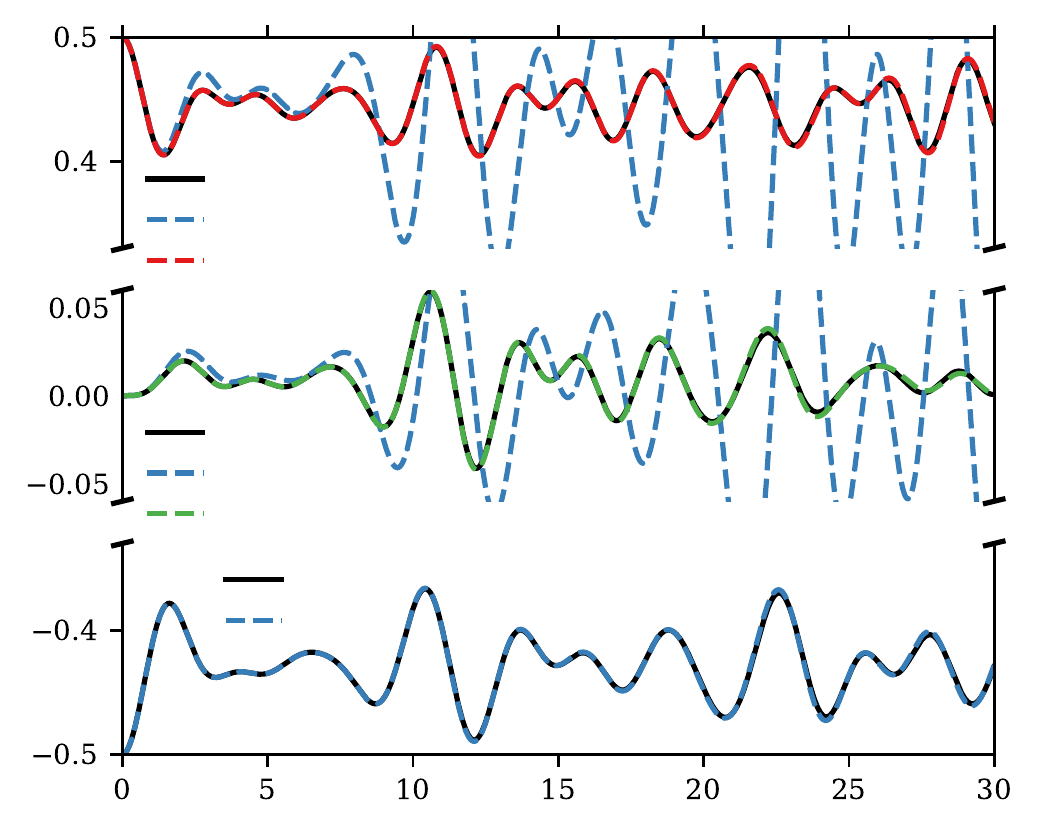}
      };
      \node[black] at (0.35,-4.2) {$t$};
      \node[black,rotate=90] at (-5.15,0.15
      7) {$\langle \mathcal{S}_l^z\rangle$};
      \node[black] at (5.9-8.65,3.1-0.65) {ED};
      \node[black] at (5.9-8.65,2.68-0.65) {$\tilde{\mathrm{T}}_4$};
      \node[black] at (7-8.65,2.25-0.65) { Particle-hole $\tilde{\mathrm{T}}_4$};
      \node[black] at (5.9-8.65,0.5-0.65) {ED};
      \node[black] at (5.9-8.65,0.05-0.65) {$\tilde{\mathrm{T}}_4$};
      \node[black] at (6.8-8.65,-0.4-0.65) {Decoupled $\tilde{\mathrm{T}}_4$};
      \node[black] at (5.9-7.85,-2.16+0.48) {ED};
      \node[black] at (5.9-7.85,-2.6+0.48) {$\tilde{\mathrm{T}}_4$};
\end{tikzpicture}
\vspace{1ex}
\caption{Benchmarking of several variants of the LKE code, including the decoupled kinetic equations \eqref{Decouple_approx}, against exact diagonalization (ED) results. As initial states we choose a fully up-polarized state $\ket{\uparrow \cdots \uparrow}$ (top), a fully down-polarized state $\ket{\downarrow \cdots \downarrow}$ (bottom), and an equal superposition of the two, $1/\sqrt{2} \ket{\downarrow \cdots \downarrow} + 1/\sqrt{2}\ket{\uparrow \cdots \uparrow}$ (centre). The bottom plot shows that, for a fully-down polarized initial state, the ``plain'' kinetic equations (as derived in the main body of the paper) with a $\tilde{T}_4$-truncation reproduce the ED results with excellent accuracy for all times shown. Indeed, according to \cref{bench_h1_final} the error at $t=30$ is of order $\Delta\langle \mathcal{S}_l^z \rangle_{\tilde{\mathrm{T}}_4} \simeq 10^{-2}$. For a fully-up polarized initial state, which corresponds to a large number of particles in the fermionic language, the plain kinetic theory with a $\tilde{T}_4$-truncation fails after relatively short times, as shown in the top plot. As explained in \cref{decoupled_dynamics}, a particle--hole transformation maps this state onto a few-particle state, and the dynamics obtained with a particle--hole-transformed kinetic theory shows excellent agreement with the ED results. Using an equal superposition of up- and down-polarised states as an initial state, neither the plain kinetic theory (nor the purely particle--hole-transformed version, which is not shown) are capable of reproducing the ED results correctly, but the decoupling approximation \eqref{Decouple_approx} achieves excellent agreement. The parameters of the Hamiltonian \eqref{Hamiltonien}--\eqref{Hpert} used for the plots are $N=10$, $\alpha=3$, $J_x=J_z=h=-1$.} \label{superposition_etats}
\end{figure}

For truncated down-polarized states $\ket{\psi^n}$ with small $n$, we have by construction that the number of fermions is small, $\bra{\psi^n} \sum_k \eta^{\dag}_{k}\eta^{\phantom{\dag}}_{k} \ket{\psi^n} \ll N/2$. As discussed in \cref{validity}, this is a requirement for our kinetic theory to provide a good approximation. For the truncated up-polarized states $\ket{\chi^n}$, in contrast, we have a large number of fermions, $\bra{\chi^n} \sum_k \eta^{\dag}_{k}\eta^{\phantom{\dag}}_{k} \ket{\chi^n} \gg N/2$, and the kinetic theory is expected to fail. However, such a state has only a small number of holes, and by applying the particle--hole transformation $\mathcal{U}$ introduced in \cref{particle_hole_trans}, it can be mapped to a state that satisfies the requirement of a small fermion number. Likewise by means of $\mathcal{U}$, the corresponding kinetic equations are obtained,
\begin{equation}
\begin{split}
id_t \bra{\chi^n} \mathcal{S}_l^z \ket{\chi^n}=&\bra{\chi^n} \mathcal{H} \mathcal{S}_l^z -\mathcal{S}_l^z \mathcal{H} \ket{\chi^n}\\
=&\bra{\psi^n} \tilde{\mathcal{H}} \mathcal{U}^\dag \mathcal{S}_l^z \mathcal{U} -\mathcal{U}^\dag \mathcal{S}_l^z \mathcal{U} \tilde{\mathcal{H}} \ket{\psi^n}
=-i \tilde{d}_t \bra{\psi^n} \mathcal{S}_l^z \ket{\psi^n}
\end{split}
\end{equation}
where $\tilde{\mathcal{H}} \coloneqq \mathcal{H}(-h,-J_x)$ is the transformed Hamiltonian and $\tilde{d}_t$ denotes the time-differential operator under $\tilde{\mathcal{H}}$. The time-evolution of a superposition
\begin{equation}
\ket{\phi^n}=  y_1 \ket{\psi^n}+ y_2 \ket{\chi^n}
\end{equation}
of truncated up- and down-polarized states is then given by
\begin{equation}\label{dynamique_superp}
d_t \bra{\phi^n} \mathcal{S}_l^z \ket{\phi^n}=D_t \bra{\psi^n} \mathcal{S}_l^z \ket{\psi^n} + 2 \myIm \left\{y_1 \overline{y}_2 \Tr \left( \left[ \mathcal{H}, \mathcal{S}_l^z \right] \ket{\psi^n} \bra{\chi^n}  \right)  \right\}
\end{equation}
where we have defined $D_t \coloneqq |y_1|^2 d_t - |y_2|^2 \tilde{d}_t$. If $\ket{\psi^n}$ and $\ket{\chi^n}$ are few-fermion vectors, the second term on the right hand side of Eq.~\eqref{dynamique_superp} is small. This is a consequence of the fact that the $p$-particle number of the terms in $\left[ \mathcal{H}, \mathcal{S}_l^z \right]$ is at most $p=4$, and hence does not couple few-fermion states to few-hole states in Fock space. Therefore, for $\alpha$ and $t$ sufficiently small, the dynamics of few-fermion and few-hole states approximately decouples,
\begin{equation} \label{Decouple_approx}
d_t \bra{\phi^n} \mathcal{S}_l^z \ket{\phi^n} \simeq D_t \bra{\psi^n} \mathcal{S}_l^z \ket{\psi^n}.
\end{equation}
On the practical side, this implies that the time-evolution of a superposition $\ket{\phi^n}$ of truncated up- and down-polarized initial state can be computed with the LKE code by making use of the original Hamiltonian $\mathcal{H}$ as well as its field-inverted counterpart $\tilde{\mathcal{H}}$.
We expect that the decoupling approximation \eqref{Decouple_approx} works well when $n$ is small, and, for fixed $n$, becomes better with increasing $N$. The excellent performance of the LKE code and the decoupling approximation is illustrated and benchmarked in Fig.~\ref{superposition_etats}.

\section{Initial conditions}

\subsection{Truncated polarized states and \texorpdfstring{$p$}{p}-particle structure}
In this subsection we sketch the main steps required in rewriting $\ket{\downarrow \cdots \downarrow}$ in the $\eta$-basis. We also comment on the link between this state and the $p$-particle structure, elaborating in particular on the role of the system size $N$ and the parameter $h/J_x$. These arguments will provide the main motivation for using small-$n$ truncated polarized states in \cref{appli_relax}.
\paragraph{Fully polarized state}
Applying the Jordan-Wigner and Fourier transformations, the fully down-polarized spin state $\ket{\downarrow \cdots \downarrow}$ is mapped onto the Fourier vacuum $\ket{0^{\mathrm{F}}}$,  defined as the only state satisfying 
\begin{equation} \label{defFockF}
\tilde{\mu}_k \ket{0^{\mathrm{F}}} =0\quad\forall k \in \mathrm{Br}.
\end{equation}
The Bogoliubov transformation $\tilde{\mu}_k \mapsto u_k \eta_k  -iv_k \eta_{-k}^{\dag}$, however, mixes creation and annihilation operators in such a way that the Bogoliubov vacuum $\ket{0}$ for which
\begin{equation} \label{defFockB}
\eta_k \ket{0} =0\quad\forall k \in \mathrm{Br},
\end{equation}
is different from $\ket{0^{\mathrm{F}}}$. In the Fock space of Bogoliubov fermions, the Fourier vacuum can be expanded as
\begin{equation}
\ket{0^{\mathrm{F}}}= \lambda_0 \ket{0}+\sum_{r=1}^N\sum_{k_1<\dotsb<k_r}\lambda_{k_1,\dotsc,k_r}  \eta_{k_1}^{\dag}\cdots\eta_{k_r}^{\dag} \ket{0},
\end{equation}
where the $\lambda_{k_1,\dotsc,k_r}$ are the coefficients we need to determine. Since the operator $u_k \eta_k - i v_k \eta^{\dag}_{-k}$ is block anti-diagonal in the decomposition $H=H_{\mathrm{even}} \oplus H_{\mathrm{odd}}$ of the Bogoliubov-Fock space, condition \eqref{defFockF} remains true for the restrictions $\tilde{\mu}_k |_{\mathrm{even}}$ and $\tilde{\mu}_k |_{\mathrm{odd}}$ of $\tilde{\mu}_k$ to the even and odd sectors, respectively. One can then prove that
\begin{equation} \label{Aprouver}
\sum_{k_1<\dotsb<k_{2q+1}} |\lambda_{k_1,\dotsc,k_{2q+1}}|^2=0\qquad\forall q \in\llbracket 0,\lfloor N/2 -1/2 \rfloor\rrbracket,
\end{equation}
i.e.\ vectors in the odd sector do not contribute to $\ket{\downarrow \cdots \downarrow}$. To prove this result, we apply \eqref{defFockF} to $\tilde{\mu}_k |_{\mathrm{odd}}$, which yields
\begin{equation} \label{une_formule}
\lambda_k u_k \ket{0}+ \sum_{q \geq 1}  \left\{ u_k \sum_{\bm{k}(q)} \Lambda_{\bm{k}(q)}^{\underline{k}} \eta_k^{\phantom{\dag}}   \mathcal{O}_{\bm{k}(q)} - i v_k \sum_{\bm{k}(q-1)}   \Lambda_{\bm{k}(q-1)}^{\cancel{-k}} \eta^{\dag}_{-k} \mathcal{O}_{\bm{k}(q-1)} \right\}   \ket{0}   =0\qquad \forall k \in \mathrm{Br},
\end{equation}
where the notation $\sum_{\bm{k}(q)}$ with $\bm{k}(q)=(k_1,\dotsc,k_{2q+1})$ denotes a summation over all $k_1<\dotsb<k_{2q+1}$. Here we have defined $\mathcal{O}_{\bm{k}(q)}:= \eta^{\dag}_{k_1}\cdots \eta^{\dag}_{k_{2q+1}}$ and have labelled by $\underline{k},~\cancel{k}$ a coefficient that contains, respectively excludes, the momentum $k$ in its definition, e.g.
\begin{equation}
\Lambda_{\bm{k}(q)}^{\underline{k}}=
\begin{cases}
\lambda_{k_1,\dotsc,k_{2q+1}}^{\phantom{\dag}} & \text{if $k \in \left\{ k_1,\dotsc,k_{2q+1} \right\}$,}\\
0 & \text{else},
\end{cases}
\end{equation}
with an analogous definition for $\Lambda_{\bm{k}(q)}^{\cancel{-k}}$.
Projecting \cref{une_formule} onto $\ket{0}$ one finds that $\lambda_k=0$ for all $k$. For $q\geq1$, we note that $\eta_k^{\phantom{\dag}} \mathcal{O}_{\bm{k}(q)}$ and $\eta_{-k}^\dag \mathcal{O}_{\bm{k}(q-1)}$ are of degree $2 q$, so that the projection on any $2q$-excited state gives a recurrence relation between some of the $\Lambda_{\bm{k}(q)}^{\underline{k}}$ and some of the $\Lambda_{\bm{k}(q-1)}^{\cancel{k}}$. For instance, for $q=1$ one obtains
\begin{equation}
\sum_{k_2<k_3} u_k \Lambda^{\underline{k}}_{k,k_2,k_3} \eta_{k_2}^\dag \eta_{k_3}^\dag \ket{0}- i v_k \sum_{k_1} \Lambda^{\cancel{-k}}_{k_1} \eta_{-k}^\dag \eta_{k_1}^\dag \ket{0}=0\qquad\forall k \in \mathrm{Br}.
\end{equation}
It then follows that, for any $k_2,k_3$ such that $-k \not\in \left\{ k_2, k_3 \right\}$, we have $\Lambda^{\underline{k}}_{k,k_2,k_3}=0$. Otherwise,
\begin{equation}
\lambda_{k,-k,k_1}=i \frac{v_k}{u_k} \lambda_{k_1}\qquad \forall k_1 \not\in \left\{-k,k \right\},
\end{equation}
and reasoning by induction leads to Eq.~\eqref{Aprouver}. Applying \eqref{defFockF} to $\mu_k |_{\mathrm{even}}$ gives similar results for all the $\lambda_{k_1,\dotsc,k_{2q}}$ coefficients, except that $\lambda_0 \neq 0$. Finally one obtains $\ket{\downarrow\cdots\downarrow}= \mathcal{G}_{\lfloor N/2 \rfloor} \ket{0}$, with
\begin{equation} \label{transf_ini_state}
\mathcal{G}_{\lfloor N/2 \rfloor}=W^{-1}\left(1+\sum_{s=1}^{\lfloor N/2 \rfloor} (-i)^s \sum_{0<k_1<\dotsb<k_s<\pi} \frac{v_{k_1}}{u_{k_1}} \cdots  \frac{v_{k_s}}{u_{k_s}} \eta_{-k_s}^{\dag}\cdots\eta_{-k_1}^{\dag}\eta_{k_1}^{\dag}\cdots\eta_{k_s}^{\dag} \right),
\end{equation}
where $W^2=1+\sum_{s=1}^{\lfloor N/2 \rfloor} L_s$ with
\begin{equation} \label{qte_Ls}
L_s=\sum_{0<k_1<\dotsb<k_s<\pi} \left( \frac{v_{k_1}}{u_{k_1}}\cdots\frac{v_{k_s}}{u_{k_s}}  \right)^2.
\end{equation}
We prefer Eq.~\eqref{transf_ini_state} over the exponential formulation $\mathcal{G}_{\lfloor N/2 \rfloor}=W^{-1}\exp\left[ i \sum_{0<k<\pi} \left( v_k/u_k \right) \eta_{-k}^\dag \eta_k^\dag \right]$ that has been used in the literature \cite{CalabreseEsslerFagotti12}, as \eqref{transf_ini_state} naturally lends itself to the definition of truncated polarized states, as discussed in the next paragraph.

\begin{figure}[tb]\centering
\begin{tikzpicture}
      \node (blabla) at (0,0){
        \includegraphics[scale=1]{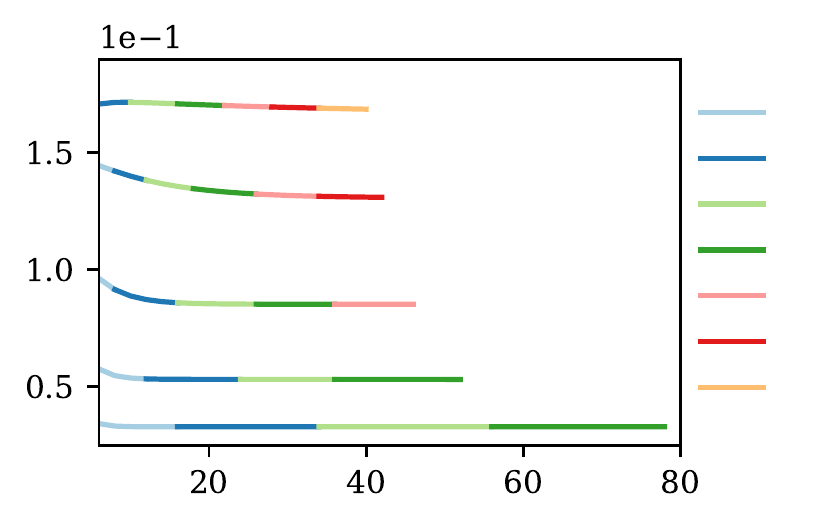}
      };
      \node[black] at (-0.15,-2.6) {$N$};
      \node[black,rotate=90] at (-4.28,0.15
      7) {$\langle \mathcal{D}\rangle$};
      \node[black,rotate=0] at (-1.25,1.83) {$h=-0.51$};
      \node[black,rotate=0] at (-1.13,0.95) {$h=-0.55$};
      \node[black,rotate=0] at (-0.8,-0.17) {$h=-0.65$};
      \node[black,rotate=0] at (-0.25,-0.95) {$h=-0.8$};
      \node[black,rotate=0] at (2,-1.4) {$h=-1$};
      
      \node[black,rotate=0] at (4.25,1.55) {$n=3$};
      \node[black,rotate=0] at (4.25,1.55-1*0.465) {$n=4$};
      \node[black,rotate=0] at (4.25,1.55-2*0.465) {$n=5$};
      \node[black,rotate=0] at (4.25,1.55-3*0.465) {$n=6$};
      \node[black,rotate=0] at (4.25,1.55-4*0.465) {$n=7$};
      \node[black,rotate=0] at (4.25,1.55-5*0.465) {$n=8$};
      \node[black,rotate=0] at (4.25,1.55-6*0.465) {$n=9$};
\end{tikzpicture}
\vspace{1ex}
\caption{Bogoliubov fermion densities $\langle \mathcal{D} \rangle$ of the fully down-polarized state $\ket{\downarrow \cdots \downarrow}$ are plotted as a function of the system size $N$ for different values of $h$ and parameter values $\alpha=4$, $J_x=J_z=-1$. For larger $N$, $\ket{\downarrow \cdots \downarrow}$ is approximated by a truncated polarized state $\ket{\psi^n}$ with $n$ large enough such that, for a given $h$, $\langle \mathcal{D} \rangle_{\psi^n}\simeq \langle \mathcal{D} \rangle_{\psi^{n-1}}$ to a precision of $10^{-3}$. For magnetic fields $h$ close to $-1/2^+$, a $\ket{\psi^n}$ with a fairly large $n$-value is required to reach the desired level of accuracy, which puts an $h$-dependent limit on the system sizes for which we were able to calculate $\langle\mathcal{D}\rangle$. From the plot we observe that (i) the criterion \eqref{few_particle_cond} for the validity of the approximations made in the LKE code does not hold for $h$ close to $-1/2$; and (ii) at fixed magnetic field, $\langle \mathcal{D} \rangle$ tends to a nonzero value when $N$ goes to the large sizes limit ($N \sim 10^2$). Note that, since the value of $n$ is not kept constant along each of the lines in the plot, such a nonzero limit is not in contradiction to \eqref{limitD}. From this limiting behaviour in combination with Eq.~\eqref{limitD} one can infer that, in order to approximate $\langle \mathcal{D} \rangle$ to a certain precision, $n$ has to increase linearly with $N$, which becomes computationally impractical for larger system sizes. Note that this analysis only concerns initial states; since $\mathcal{D}$ does not commute with $\mathcal{H}$, its expectation value changes with time, which may (and it practice does) lead to a violation of the criterion \eqref{few_particle_cond} at later times.} \label{polarized_state_densities}
\end{figure}
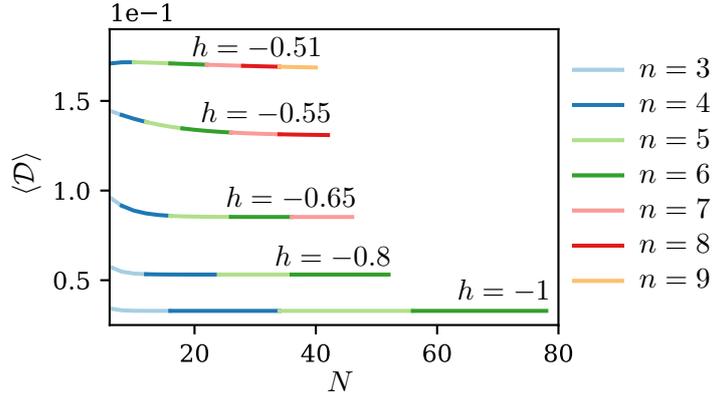

\paragraph{Truncated polarized states: properties and limitations}\label{restriction} 
The $n$th truncated polarized state $\ket{\psi^n}$, as defined in Eq.~\eqref{truncstates}, is obtained by truncating the first sum in \eqref{transf_ini_state} at the index $s=n$. Such a truncation preserves the reflection symmetry in momentum space, and therefore allows us to make use of the symmetry restrictions discussed in \cref{symmetries}. Also, the numerical computation of the vector $\ket{\psi^n}$ scales like $N^n$, whereas that of the fully polarised state scales like $\lfloor N/2 \rfloor !$. Finally, truncated polarized states come with the additional advantage of allowing for a systematic tuning of the particle density such that, by making $n$ sufficiently small, the validity of the LKE code can be ensured. This can be useful for small magnetic fields $h$, where $\bigl\langle \sum_k \eta_k^\dag \eta_k^{\phantom{\dag}} \bigr\rangle \ll N/2$ in general does not hold for a (non-truncated) polarized state, as illustrated in \cref{polarized_state_densities}.

A straightforward calculation shows that the expectation value of the fermionic particle density \eqref{fermiondensity} with respect to any truncated polarized state is given by
\begin{equation}\label{D_psi_N}
\langle \mathcal{D} \rangle_{\psi^n}=\frac{1}{N}\frac{\sum_{s=1}^{n} 2s L_s}{1+\sum_{s=1}^{n} L_s}.
\end{equation}
For the moment let us assume that, for sufficiently large $N$, $n$ being fixed,
\begin{equation}\label{asymptoticsLs}
L_1(N) \ll L_2(N) \ll  ... \ll L_s(N),\qquad \forall s\in\llbracket 1,n\rrbracket,
%L_s(N)=\mathcal{O}(N^s)\qquad \forall s\in\llbracket 1,n\rrbracket,
\end{equation}
which will be justified towards the end of the section. Then it follows from \eqref{D_psi_N} and \eqref{asymptoticsLs} that
\begin{equation}\label{limitD}
\langle \mathcal{D} \rangle_{\psi^n} \lesssim \frac{2n}{N}
\end{equation}
for sufficiently large $N$. From the upper bound \eqref{limitD}, the criterion \eqref{few_particle_cond} for the validity of the $p$-particle truncation for a truncated polarized initial state $\ket{\psi^n}$ immediately follows for $p \gtrsim n$. Moreover, since $\ket{\psi^n}=\ket{\downarrow\cdots\downarrow}$ for $n=\lfloor N/2\rfloor$, there must exist a smallest-possible $\nu$ for which, for a given system size $N$,  $\langle \mathcal{D} \rangle_{\psi^\nu} \simeq \langle \mathcal{D} \rangle_{\downarrow}$ to a desired level of accuracy. It seems reasonable to assume that $\ket{\psi^\nu}$ is then a good approximation of the corresponding fully-polarized vector.

We chose to present in \cref{appli_relax} LKE results only for $\ket{\psi^1}$, where \eqref{few_particle_cond} holds by construction, independently of the choice of $h/J_x$. $\ket{\psi^1}$ is not necessarily a good approximation of the fully polarized state, as for instance in \cref{CampagneN120} for parameter values $N=120$ and $h=-0.51$. While also in this case $\ket{\psi^1}$ is a perfectly legitimate choice as an initial state in the LKE code, it does not have a simple (approximate) representation in the spin picture, and the physical relevance of such an initial state is unclear.

In the remainder of this section we complete the above reasoning by providing a justification of the asymptotic property \eqref{asymptoticsLs}. We start by showing that $L_1$ grows linearly with $N$ asymptotically in the large-$N$ limit. From the definition \eqref{qte_Ls} of $L_s$, together with the expressions of the Bogoliubov coefficients $u_k$ and $v_k$ derived in \cref{Bogoliubov}, it follows that
\begin{equation}\label{L1}
L_1(N)=\sum_{\theta \in \text{Br}} g \circ f_\kappa(\theta),
\end{equation}
where
\begin{equation}
f_\kappa(\theta)=\left(\frac{\sin \theta}{\kappa+ \cos \theta}\right)^2,\qquad
g(y)=\frac{1}{2}\frac{\sqrt{1+y}-1}{\sqrt{1+y}+1},
\end{equation}
and $\kappa=2h/J_x$ is the order parameter. For $\kappa>1$, each term in the sum of \eqref{L1} is positive and smaller than one, which implies $L_1(N)\leq N$. To also prove a lower bound on $L_1$, we define, for a fixed $0<\epsilon \ll \pi/4$, the interval
\begin{equation}
I=\left[- \pi + \epsilon, - \epsilon \right]\cup \left[ \epsilon, \pi - \epsilon \right],
\end{equation}
chosen such that, for a fixed $M>1$, for all $\kappa\in \left]1 , M \right]$ and $\theta\in I$ we have
\begin{equation}
f_{\kappa}(\theta) \geq A=\mathrm{min} \left[f_{\mathrm{M}}(\epsilon),f_{\mathrm{M}}(\pi - \epsilon) \right]>0.
\end{equation}
Then it follows that
\begin{flalign}\label{L1inequality}
L_1(N)=\sum_{\theta\in\text{Br}} g(f_\kappa(\theta))& \geq \sum_{\theta \in \text{Br} \cap I} g(f_\kappa(\theta)) \geq g(A)  \sum_{\theta \in \text{Br} \cap  I} 1 \geq \frac{Ng(A)}{2},
\end{flalign}
where the first inequality follows from positivity of $g$, and the second from the fact that $g$ is an increasing function on its domain. Note that $g(A)>0$ because, by construction, $A>0$. The third inequality in \eqref{L1inequality} is then valid except for very small $N$, where the discreteness of the Brillouin zone may spoil it.
Taking the upper and the lower bound together, it follows that $L_1(N)=\Theta(N)$. Along similar lines one can show that $L_s(N)=\Theta(N^s)$, which implies \eqref{asymptoticsLs}.

\subsection{Initial expectation values}
 \label{Some_initial_analytics}
In this section we gather expressions of the energy density $\nu=\langle \mathcal{H} \rangle/N$ and the spin observable $\langle \mathcal{S}_l^z \rangle$ for several specific states: $\ket{0}$, $\ket{\psi^1}$ and $\ket{\downarrow \cdots \downarrow}$, which are referred to in the main text, as well as their particle--hole counterparts $\ket{N}$, $\ket{\chi^1}$ and $\ket{\uparrow \cdots \uparrow}$, which are required in \cref{high_temp_dynamics}.

\subsubsection{Energy density} \label{energy_densit}
We obtain $\nu_0= \langle \mathcal{H}_0 \rangle / N$ for the Bogoliubov vacuum $\ket{\psi^0}=\ket{0}$, and $\nu_{\lfloor N/2 \rfloor}=-h/2 + J_z \zeta /8$ for the fully $z$-polarised state $\ket{\psi^{\lfloor N/2 \rfloor}} = \ket{\downarrow ... \downarrow}$. For the truncated polarised state $\ket{\psi^1}$ one finds
\begin{equation} \label{nu_1_down}
\begin{split}
\nu_1 &\equiv\bra{\psi^1}\mathcal{H}\ket{\psi^1}\\
&= \nu_0 + \frac{2}{N W_1^2} \sum_{j>0} \left\{ 2i \frac{v_j}{u_j} A_{\mbox{\tiny I}}(j)+ \left( \frac{v_j}{u_j} \right)^2 A_{\mbox{\tiny II}}(j)
+ \sum_{l>0} \frac{v_j}{u_j}\frac{v_l}{u_l} \left( B_{\mbox{\tiny III}}(l,l,j,j) - B_{\mbox{\tiny III}}(-l,-l,j,j) \right)  \right\}.
\end{split}
\end{equation}
The particle--hole counterpart $\overline{\nu}_1 := \bra{\chi^1} \mathcal{H}\ket{\chi^1}/N$ is identical to \eqref{nu_1_down} under magnetic field reversal $h\to-h$. For $N>8$ one finds $\bra{\chi^1}\mathcal{H} \ket{\psi^1}=0$, and hence the energy density of the superposition $\ket{\phi^1}=y_1 \ket{\psi^1}+ y_2 \ket{\chi^1}$ simplifies to
\begin{equation}
\nu_{\phi^1}\equiv\bra{\phi^1}\mathcal{H}\ket{\phi^1}=|y_1|^2 \nu_1 +|y_2|^2 \overline{\nu}_1,
\end{equation}
where $y_1$ and $y_2$ are complex coefficients normalized such that $|y_1|^2 +|y_2|^2 =1$. As a rule of thumb,
\begin{equation} \label{energy_dens_psi1_approx}
\nu_1  \simeq - \sum \epsilon_k /2 N,
\end{equation}
so that for $h<0$ it follows with \eqref{dispersion_relation} that $\nu_1 >0$ and $\overline{\nu}_1 < 0$. It is therefore possible to choose $y_1,y_2$ such that $\nu_{\phi^1}\simeq 0$, which turns out to be useful in \cref{particle_hole_trans} as a way of constructing initial states in the regime of small energy densities, which, according to \cref{taylor1}, correspond to small inverse temperatures. To calculate the energy densities related to other truncated polarized states, we rely on the LKE code, which gives numerically exact initial energy densities for all truncations that contain $\tilde{\mathrm{T}}_{4}$ as a subset. Finally, for the state $\ket{\phi^{\lfloor N/2 \rfloor}}$, which is used in \cref{superposition_etats}, one can check that
\begin{equation} \label{energydens_initial_superp}
\upsilon_{\phi^{\lfloor N/2 \rfloor}}=h \, \frac{\left|y_2 \right|^2-\left|y_1 \right|^2}{2}  + \frac{J_{z} \zeta}{8},
\end{equation}
and thus it is once more possible to adjust $y_1$ and $y_2$ such that $\upsilon_{\phi^{\lfloor N/2 \rfloor}}=0$.

\subsubsection{Spin observable} \label{spin_initial_mean_val}
We obtain $\langle \mathcal{S}_l^z \rangle = \pm 1/2$ for a fully $z$-polarized initial state, and
\begin{equation} \label{Slz_psi0}
\langle \mathcal{S}_l^z \rangle_0 =\Gamma_N =-\langle \mathcal{S}_l^z \rangle_N
\end{equation}
for the Bogoliubov vacuum $\ket{0}$ and the anti-vacuum $\ket{N}$, with $\Gamma_N$ as defined in \eqref{Gamma_N}. For $\ket{\psi^1}$ we find
\begin{equation} \label{Slz_psi1}
\langle \mathcal{S}_l^z \rangle_1 = \langle \mathcal{S}_l^z \rangle_0 - \frac{2}{W_1^2 N} \sum_{k >0}  \frac{v_k}{u_k}  \left(2 Y_{kk} - \frac{v_k}{u_k} X_{kk}   \right),
\end{equation}
where the second term on the right-hand side is a correction to $\langle \mathcal{S}_l^z \rangle_0$ of order $O(1/N)$. Indeed,
\begin{equation} \label{bound1}
\left| \langle \mathcal{S}_l^z \rangle_1 - \langle \mathcal{S}_l^z \rangle_0 \right| = \frac{2}{W_1^2 N} \sum_{k >0} v_k^2 \left( 1+  \frac{v_k^2}{u_k^2}  \right) \leq 2/N.
\end{equation}
Moreover, \eqref{Gamma_N} [and hence \eqref{Slz_psi0}] can be regarded as a Riemann sum, and one can take the continuum limit $N \to \infty$. At finite $N$, the error made in the substitution by an integral is bounded by a term of order $O(1/N)$, and using \eqref{bound1} we obtain
\begin{equation}
\langle \mathcal{S}_l^z \rangle_1= -\frac{1}{2} \int_0^1  dx / \sqrt{f_{\kappa}( \pi x)+1} + O(1/N).
\end{equation}
For $J_x=-1$ this implies that
\begin{equation} \label{bounds_Slz}
\lim_{h \to - \infty} \lim_{N \to + \infty} \langle \mathcal{S}_l^z \rangle_1 = -1/2, \qquad  \lim_{h \to - 1/2} \lim_{N \to + \infty} \langle \mathcal{S}_l^z \rangle_1 = -1/\pi.
\end{equation}
Since $h \mapsto \lim_{N \to + \infty} \langle \mathcal{S}_l^z \rangle_1 (h)$ is strictly increasing on $]{-\infty}, -1/2]$, 
Eq.~\eqref{bounds_Slz} provides an analytical justification of the numerically observed initial values of $\langle \mathcal{S}_l^z \rangle$ in Figs.~\ref{quadratic_truncation} and \ref{CampagneN120}. In \cref{CampagneN120} for instance where $N=120$, the initial values predicted by \eqref{bounds_Slz} are $\langle \mathcal{S}_l^z \rangle_1 \simeq -0.33$ for $h=-0.51$ and $\langle \mathcal{S}_l^z \rangle_1 \simeq -0.46$ for $h=-1$, at order zero in $1/N$. One can then check that, with the exact expression \eqref{Slz_psi1}, the error is indeed smaller than $10^{-2}$.

\bibliographystyle{naturemag}
\bibliography{MK}

\end{document}